\documentclass[10pt]{iopart}
\pdfoutput=1
\usepackage{iopams} 
\usepackage{graphicx}
\usepackage{epsfig}
\usepackage{float}

\def\fun#1#2{\lower3.6pt\vbox{\baselineskip0pt\lineskip.9pt
\ialign{$\mathsurround=0pt#1\hfil##\hfil$\crcr#2\crcr\sim\crcr}}}

\def\mpc{\,\rm Mpc}

\begin{document}

\title[Photo-z optimizaton]{Photo-z optimization for measurements of the BAO radial scale}
\author{Daniel Roig$^1$, Licia Verde$^{3,1}$, Jordi Miralda-Escud\'e$^{3,2,1}$,
Raul Jimenez$^{3,1}$, Carlos Pe\~na-Garay$^{4}$\\ 
$^1$ {\it Institute of Space Sciences (CSIC-IEEC), Fac. Ci\`encies, Campus UAB, Bellaterra, Spain.} \\
$^2$ {\it Institut of Ci\`encies del Cosmos, Universitat de Barcelona, Barcelona, Spain.} \\
$^3$ {\it ICREA, Barcelona, Spain} \\
$^4$ {\it Instituto de F\'isica Corpuscular (CSIC-UVEG), Val\`encia, Spain}\\}

\begin{abstract}

Baryon Acoustic Oscillations (BAO) in the radial direction offer a method to
directly measure the Universe expansion history, and to set limits to space
curvature when combined to the angular BAO signal. In addition to spectroscopic
surveys, radial BAO might be measured from accurate enough photometric
redshifts obtained with narrow-band filters. We explore the requirements for a
photometric survey using Luminous Red Galaxies (LRG) to competitively measure
the radial BAO signal and discuss the possible systematic errors of this
approach. If LRG were a highly homogeneous population, we show that the photo-z
accuracy would not substantially improve by increasing the number of
filters beyond $\sim 10$, except for a small fraction of the sources
detected at high signal-to-noise, and broad-band filters would suffice
to achieve the target $\sigma_z = 0.003 (1+z)$ for measuring radial BAO.
Using the LRG spectra obtained from SDSS, we find that the spectral variability 
of LRG substantially worsens the achievable photometric redshift
errors, and that the optimal system consists of $\sim$ 30 filters of width
$\Delta \lambda / \lambda \sim 0.02$. A $S/N > 20$ is generally necessary at
the filters on the red side of the $H\alpha$ break to reach the target
photometric accuracy. We estimate that a 5-year survey in a dedicated telescope 
with etendue in excess of 60 ${\rm m}^2\, {\rm deg}^2$ would be necessary to obtain 
a high enough density of galaxies to measure radial BAO with sufficiently low shot 
noise up to $z= 0.85$. We conclude that spectroscopic surveys have a superior performance
than photometric ones for measuring BAO in the radial direction.

\end{abstract}  

\section{Introduction}

An important observable to constrain the nature of dark energy is the
Hubble parameter $H(z)$ \cite{SVJ05}, since it constitutes a more direct probe to the
dark energy equation of state than the angular diameter distance
$d_a(z)$ or the luminosity distance $d_L(z)$, which depend on an
integral of $H(z)$. Einstein's equations imply that a homogeneous and
isotropic universe, which is described by the FRW metric, that is
composed of matter and dark energy with equation of state
$p_Q = w_Q(z) \rho_Q$ expands according to
\begin{eqnarray}
\frac{H(z)}{H_0} = \left[\rho_T(z)\over \rho_T(0) \right]^{1/2} = \\ \nonumber
\left [ \Omega_M (1+z)^3+ \Omega_k(1+z)^2+\Omega_Q\exp\! \left ( 
3 \int_0^z \frac{1+w_Q(z')}{1+z'} dz' \right ) \right ]^{1/2}\!,
\end{eqnarray}
where the subscripts $Q$, $k$, $M$ and $T$ refer to dark energy, space
curvature, matter, and total energy density, respectively. In the
absence of space curvature, the quantities $d_A(z)$ and $d_L(z)$ are
related to $H(z)$ via $d_A(z)(1+z)=d_L(z)/(1+z) = \int_0^zdz'/H(z')$. 

The acoustic oscillations in the photon-baryon plasma, observed as
acoustic peaks in the CMB power spectrum, are also imprinted in the
matter distribution at the scale of the sound horizon at the radiation
drag epoch, when baryons were released from the photon pressure. The
mass distribution is traced by the distribution of galaxies, and the
Baryon Acoustic Oscillations (BAO) can be observed as a peak in the
galaxy correlation function or as a series of harmonic oscillations in
the galaxy power spectrum. The sound horizon at the radiation drag epoch
can be computed very accurately from CMB observations
(e.g., $r_s=153.2\pm 2.0$ Mpc from WMAP5, \cite{wmap5}), so it provides
a natural standard ruler. In fact, the galaxy power spectrum can be used
to measure both the angular diameter distance through the clustering  
perpendicular to the line-of-sight, and the expansion rate $H(z)$
through the clustering along the line-of-sight. 
Therefore, BAO measurements test the relation between
$H(z)$ and $d_A(z)$, providing constraints on dark energy and a limit to
space curvature (e.g.,
\cite{Polarsky05,Huangetal06,Clarcksonetal07,Barenboimetal08} ).
The BAO technique is being considered a powerful probe
to the nature of dark energy \cite{sdss,2dF,Percival,Percival07} because of its
potential to provide a standard ruler at different redshifts and its robustness to systematic effects.

For an ideal galaxy survey, \cite{SeoEisenstein07} have shown that a
volume of $1$ (Gpc/h)$^3$ at low
redshift can constrain $H_0$ to the 7 \% level. Forthcoming surveys with
larger volume are expected to reach the statistical power to constrain
$H(z)$ at the \% level. There are a number of requirements that these
surveys need to satisfy for measuring BAO: covering a large survey
volume, modeling the effects of galaxy bias and non-linearity,
characterizing the covariance between different modes, evaluating the
galaxy selection function to sufficient accuracy, reducing photometric
calibration errors to low enough levels to avoid contamination of the
BAO signal, etc. Measuring the BAO scale in the radial direction demands
in addition that galaxy redshifts are measured to a sufficiently high
accuracy, $\sigma_z$, to avoid an excessive smoothing of the BAO peak,
which has an intrinsic width $\delta r \sim 10$ Mpc, i.e.,
$\delta z \leq \delta r H(z)/c$.
As shown by \cite{SeoEisen03}, a redshift accuracy
$\sigma_z < 0.003(1+z)$ is required to avoid substantial loss of
accuracy of the $H(z)$ measurement from a given survey volume.
Spectroscopic surveys usually yield a redshift accuracy much higher
than this minimum requirement.

An alternative approach to measure the large number of redshifts
required for BAO detection are photometric redshifts from imaging
surveys. Broad-band photometry with $\sim 6$ filters usually reaches
only to $\sigma_ z / (1+z) \sim 0.03$ for the general population (with
red galaxies having slightly smaller photo-z errors than blue ones),
insufficient for measuring radial BAO. However, as the Combo-17 survey
\cite{combo17} has demonstrated, the galaxy photo-z accuracy can be
improved by using a larger number of narrower filters. 
Photometric surveys can cover a large area of the sky faster than a
spectroscopic survey and reach a higher number density of observed
objects. We are therefore motivated to investigate the requirements
for a photometric survey with medium to narrow bands to deliver
interesting BAO measurements, and the optimization of the number of
filters. Previous work has already explored this issue
(\cite{Bridle, SeoEisen03,Dahlen,Benitez}). Here we concentrate specifically
on the impact of the number of bands, signal-to-noise, and
non-uniformity of the galaxy sample.
We conclude with several considerations on the systematic effects that
the photometric approach entails.

In our investigation we use both synthetic stellar population models and
Sloan Digital Sky Survey (SDSS)-DR6 spectra of Luminous Red Galaxies. 
We concentrate on LRG because they have several properties that
make them particularly useful for BAO surveys: they are a fairly
homogeneous population, the form of their spectra makes them
particularly suitable for good photo-z determinations, and their high
luminosity facilitates reaching a high enough signal-to-noise up to high
redshifts. For these reasons, LRG are the target of choice for $z<2$
BAO surveys (at higher redshift the Ly$\alpha$ forest probably provides
the best method for measuring BAO; see \cite{McDonaldEisenstein}). 
The main conclusion we reach is that a spectroscopic survey is superior
to narrow-band photometric surveys for measuring the radial BAO signal. We also show
that if it were possible to find a very homogeneous population of LRG
(with spectra closely matched by a single spectral template), then
the photometric redshift accuracy would not substantially increase with
the number of filters used beyond a total number of $\sim 10$ at a fixed
total exposure time. This is not the case for galaxies measured at high
signal-to-noise, for which a larger number of filters is optimal, but
this high signal-to-noise cannot be achieved for a large enough number
of objects in a way that is competitive with the spectroscopic approach.
In reality, however, the
variability of realistic galaxy spectra worsens the photometric redshift
accuracy, making it optimal to increase the number of filters to
$\sim 30$ and requiring a higher signal-to-noise per filter to reach the
desired redshift accuracy. These results are generally in good agreement
with those of \cite{Dahlen}.

This paper is organized as follows: in \S~2 we review the requirements
for a survey to measure radial BAO. The modeling of the LRG population
is described in \S~3, and in \S~4 we describe our fiducial survey model.
The results are presented in \S~5, where we analyze in detail the
photo-z accuracy as a function of the number of filters, galaxy
luminosity and redshift, first for ideal galaxies that match the
templates precisely and then for real galaxies with SDSS spectra.
Discussion and conclusions are presented in \S~6 and \S~7. Throughout this paper, we use a
cosmological model with $H_0 = 70$ km s$^{-1}$  Mpc$^{-1}$,
$\Omega_m=0.3$, $\Omega_{\Lambda}=0.7$.
Readers who want to quickly see the main conclusions or our study may
wish to go directly to Figure 14. This shows the number density of LRG in
several redshift bins with photometric redshift better than $0.003(1+z)$
as a function of the etendue times the exposure time of a survey.  
It also shows the number density required to reach $nP=1$, necessary to make shot
noise subdominant in the Fourier modes near the line-of-sight useful for
measuring BAO.

\section{Spectroscopy vs photometry and target photometric requirements}

\begin{figure}
\includegraphics[width=\textwidth, angle=0]{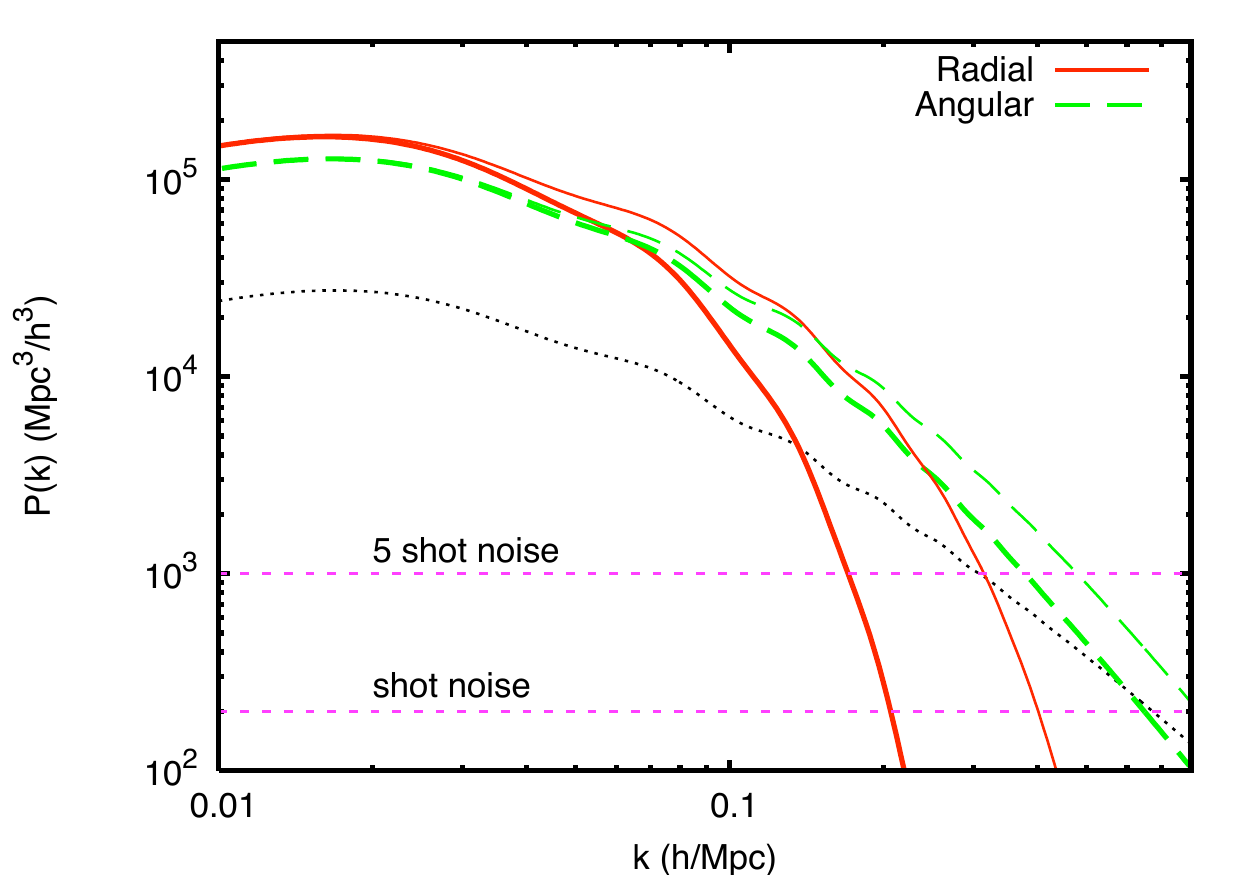}
\caption{Galaxy power spectrum in the LCDM model. {\it Dotted line:}
Present--day, real--space, linear matter power spectrum.
{\it Solid lines:} Redshift space galaxy power spectrum on the
line-of-sight, for a scale independent bias $b=2$ and redshift error
$\sigma_z=0.003$ ({\it thick line}) or assuming typical
small--scale redshift space distortions only ({\it thin line}).
{\it Long dashed lines:} Angle-averaged power spectrum, also including
redshift errors ({\it thick}) or not {\it thin}. The shot-noise
contribution level for a galaxy density of $5\times 10^{-3} h^3 \mpc^{-3}$
is shown by the short-dashed horizontal lines for reference.}
\label{fig:pk}
\end{figure}

Future BAO surveys need to: a) cover large volumes of the universe
sampling the acoustic scale $r_s = 153.2\pm 2$ Mpc with a density of
galaxies high enough to make shot-noise subdominant, and b) not degrade
with redshift errors the line-of-sight information that yields a
measurement of $H(z)$.
The latter condition implies, quantitatively, that the statistical
photo-z errors need to be smaller than $\sigma_z=0.003 (1+z)$
\cite{SeoEisen03}.
In Figure~\ref{fig:pk} we show the present--day (z=0) linear matter power spectrum 
({\it dotted line}) and the corresponding line-of-sight redshift-space
power spectrum ({\it thin solid line}), assuming a scale-independent
bias $b=2$, a $\beta$ parameter $\beta=1/b\, d\ln \delta/d\ln a\sim
\Omega_m^{0.6}/b \simeq 0.24$, and assuming
an intrinsic galaxy velocity dispersion of $420$ km/s, corresponding to
a pairwise velocity dispersion $\sigma_p=600$km/s, typical of non-linear
redshift-space distortions. The line-of-sight power spectrum is also
shown with Gaussian photo-z errors of $\sigma_z = 0.003 (1+z)$ added 
({\it thick solid lines}).
The effects of biasing, linear redshift-space distortions
\cite{kaiser}, and the combined dispersion from photo-z errors and
non-linear redshift-space distortions can be modelled as:
\begin{equation}
P_g(k,\mu)=b^2(1+\mu^2\beta)^2P(k) D(k\sigma_z \mu) ~,
\label{eq:Pk}
\end{equation}
where $\mu=\cos \theta$, and $\theta$ is the angle with respect  
to the line-of-sight. In  equation~\ref{eq:Pk}  the ``Kaiser factor" is strictly valid only for  
linear, large-scale redshift-space distortions, but it will be  
sufficient for our main purpose of illustrating the requirements on
the redshift accuracy and the shot-noise that need to be reached.
The function $D(k \sigma_z \mu)$ describes the small-scale smearing  
of power. It is often parameterized by a Gaussian,
$D(k \sigma_z \mu)= \exp [-(k\sigma_z\mu)^2]$, or by a Lorentzian,
$D(k \sigma_z \mu)= [1+ (k \sigma_z \mu)^2 ]^{-1}$.
If $x \equiv k \sigma_z \mu <<1$, the two descriptions are equivalent
since $\exp(-x^2)\simeq (1+x^2)^{-1}$.
The effect of small-scale nonlinear redshift-space distortions is
approximated as a contribution to $\sigma_z$ (subdominant in this
case compared to redshift errors). Assuming $k\sigma_z \mu \ll 1$ and
using the Lorentzian form for $D(k\sigma_z \mu)$, the angle-averaged
(monopole) power spectrum can be calculated analytically with the
following result:
\begin{eqnarray}
\int_{-1}^{1}P_g(k,\mu) d\mu \simeq \\ \nonumber
\!\!\!\!\!\!\!\!\!\!\!\!\!\!\!\!b^2 P(k)\left[ \frac{(\beta^2+6\beta)}{3 k^2\sigma_z^2} -
\frac{\beta^4}{k^4\sigma_z^4} + 
\left( \frac{1}{k\sigma_z} - \frac{\beta}{\sqrt{8} k^3 \sigma_z^3} +
\frac{\beta^2}{\sqrt{2} k^5\sigma_z^5} \right)\, {\rm arctan}(k\sigma_z)
\right] ~.
\end{eqnarray}
This monopole term is shown as the long-dashed line in Figure~\ref{fig:pk}.
The Figure also shows the shot noise contribution ({\it short-dashed
lines}) for a galaxy density of $\bar{n}=5\times 10^{-3}$, and the level  
at which $\bar{n}P=5$ and $\bar{n}P=1$ , which is where the shot noise 
contribution is  negligible: in the line-of-sight, shot noise starts becoming  
comparable to the $P(k)$ signal at larger scales than for the
angle-averaged $P(k)$.

Let us consider the following two cases: a photometric 
survey with target photo-z errors $\sigma_z \sim 0.003$ ({\it thick 
lines}), and a spectroscopic survey ({\it thin lines}, 
 corresponding to unavoidable non-linear velocities). The sampling
variance error on $P(k,\mu)$ scales as
$\sigma_P/P = 1/\sqrt{2\pi k^2\Delta k \Delta \mu  
V_{eff}/(2\pi)^3}$, where $V_{eff}=\left[\bar{n}P(k,\mu) / (\bar{n} 
P(k,\mu)+1 )\right]^2 V_{survey}$, and the accuracy at which the  
acoustic scale can be measured is directly proportional to $\sigma_P/P$.
Thus, if the two surveys have the same number density of galaxies, and
the spectroscopic survey reaches $nP=5$ along the line of sight at
$k=0.32 h/\mpc $, then the photometric survey reaches $nP=5$ at $k=0.16 h/\mpc$.
Note that the number of independent modes is roughly proportional to
$k_{max}^3$, and so the spectroscopic survey could obtain a much
better constraint on the BAO scale from measuring the power on many more
modes.
 \footnote {Since the relative error on the power spectrum $\sigma_P/P$ does not
depend on redshift, the above considerations are valid at any redshift,
with the caveat that the non-linearities decrease with $z$ but the
photo-z errors do not.}
In order to be competitive with a spectroscopic survey, a photometric
survey would need to achieve a much higher galaxy density. If the survey
volume were to be the same for the two surveys,
then at $k=0.2 h/\mpc$ a photometric survey would need a galaxy
number density $\exp(-k^2\sigma_z^2)\sim 25$ times higher than a
spectroscopic survey to achieve the same $\bar{n} P$ along the line
of sight direction.
The angle-averaged quantity is, of course, less sensitive to the
smearing along the line of sight, and the same is true for all
orientations where $\mu <1$. As shown by \cite{SeoEisen03}, at z$\sim$1
a redshift error of 0.3\% degrades the error on $H(z)$ by a factor
$\sim$ 2, demanding therefore a survey with 4 times the volume of a
spectroscopic survey to match its performance.
As most forthcoming spectroscopic
surveys will cover more than a quarter of the available sky (i.e.,
more than a quarter of the 30000 square degrees outside the galactic
plane), the photometric approach might only be advantageous if it
could reach higher redshifts. 

The above considerations indicate that a spectroscopic survey is the
favored option unless a much larger fraction of the sky can be covered  with
a photometric survey and  with an extremely high object density;  
this is equivalent to imposing the requirement of $\sigma_z<0.3$\%  
down to fainter magnitudes. Below, we calculate if such a survey is  
possible. For this we  concentrate on LRG, which  are bright and have very  
homogeneous spectra, and we study the dependence of the results on  
intrinsic galaxy variability, luminosity, etc.

\section{Models for the Population of Luminous Red Galaxies}

We start by calculating the number density of LRG  that can be
observed at each luminosity and redshift. We use the \cite{Brown}
luminosity function of LRG and adopt their model of a Shechter
luminosity function slope of $\alpha = -0.5$ (see Table 6 in \cite{Brown}).
To model the spectral distribution of galaxies, we use the first five
templates presented in \cite{Niemack} \footnote{http://www.ice.csic.es/personal/jimenez/PHOTOZ/}. These
are empirical templates computed using the SPEED \cite{Jimenez+04} and CB08 (Charlot, private communication) models with solar
metallicity and with star formation histories that
are representative of the observed distribution in red galaxies over the
stellar mass range $10^9$ to $10^{12} M_{\odot}$ extracted from
\cite{Panter07}. The spectral templates correspond to the
first five curves of Figure 4 in \cite{Panter07}. The dust reddening is performed using the method described in \cite{Panter07}.

\begin{figure}
\includegraphics[width=\textwidth, angle=0]{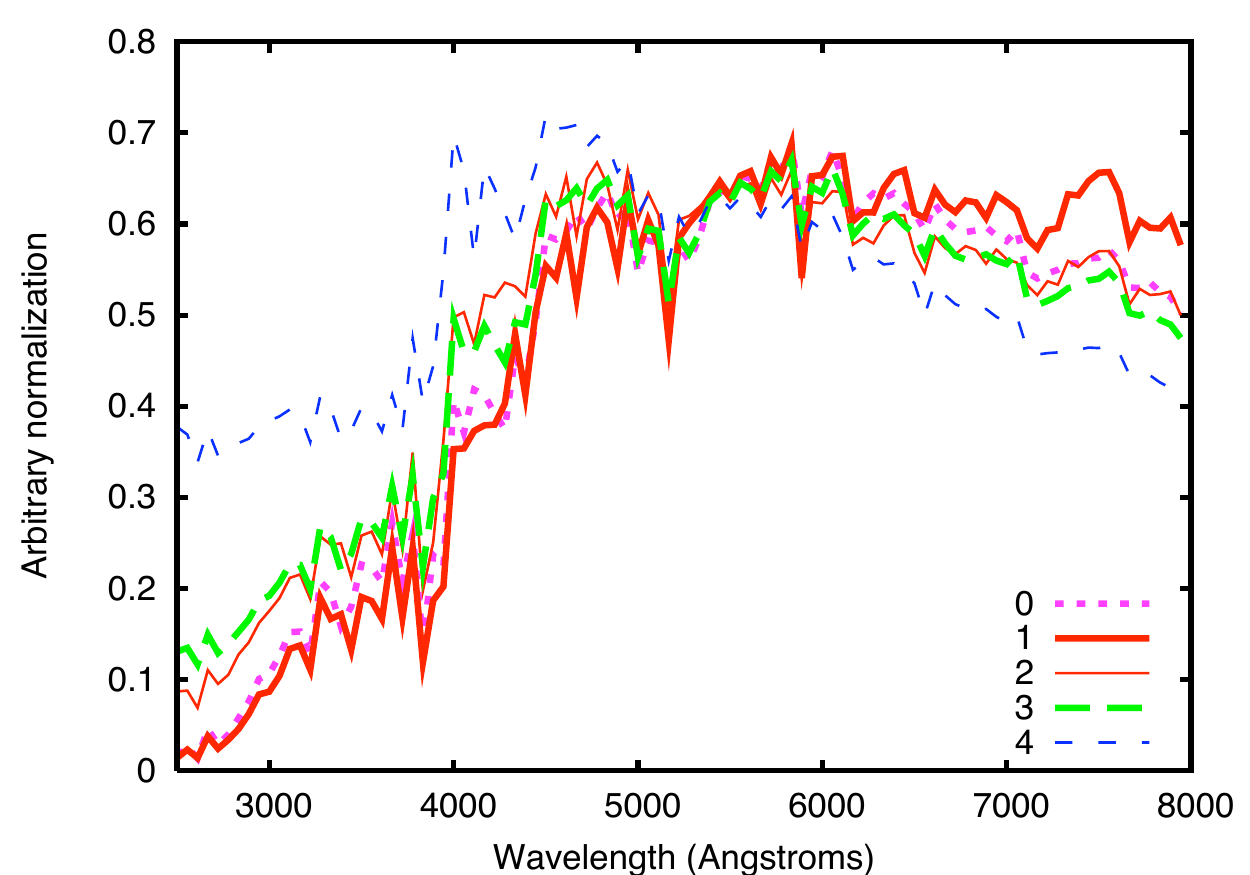}
\caption{Spectral energy distribution of the five galaxy templates used
in this paper, labelled from 0 to 4, from \cite{Niemack}. For displaying purposes, in these figures a 20 \AA\, smoothing has been applied to the lines.} 
\label{fig:seds}
\end{figure}

\begin{figure}
\includegraphics[width=\textwidth, angle=0]{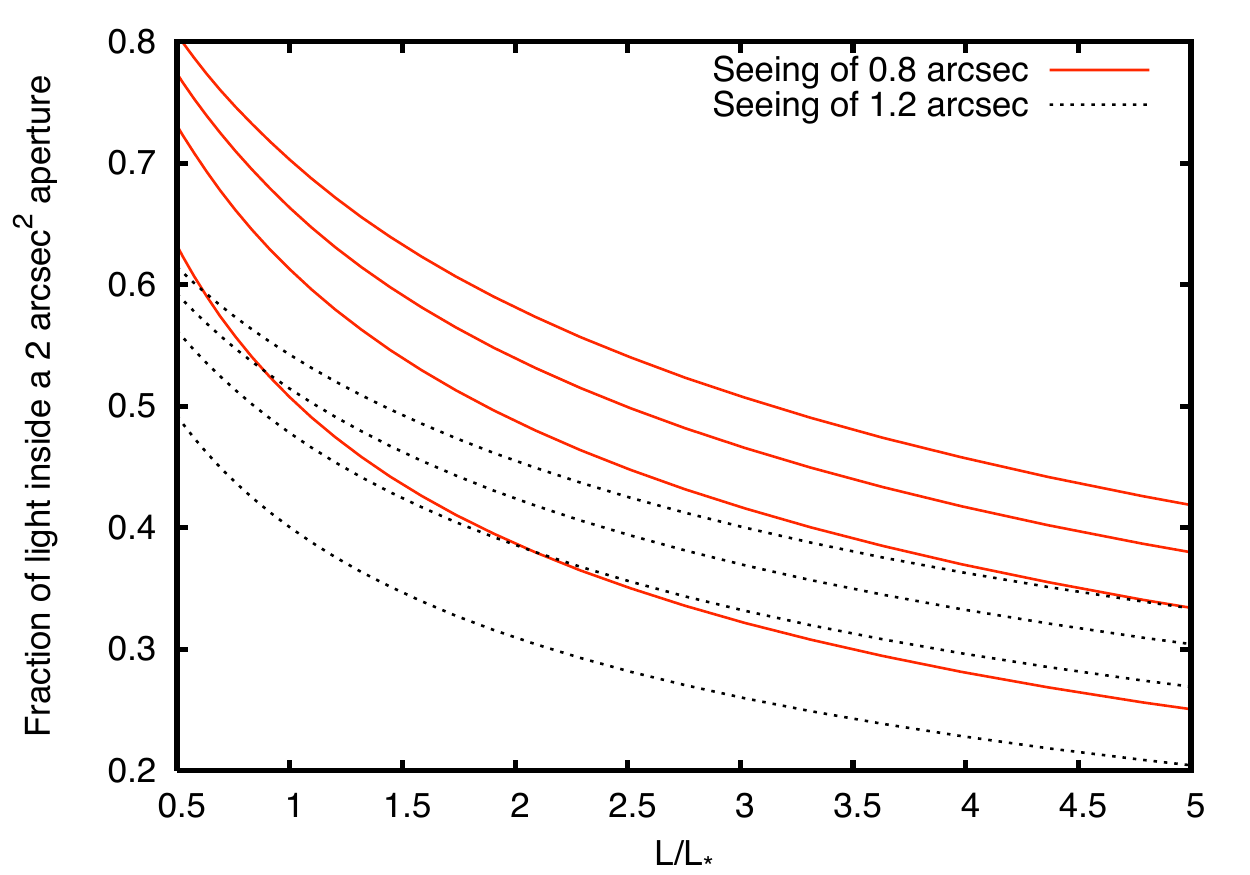}
\caption{Fraction of light of a LRG within a 2 arcsec$^2$ aperture as a
function of luminosity for four different values of redshift,
$z=0.3$, 0.5, 0.7, 0.9 from bottom to top, and two different values of
the seeing, 0.8 arcsec ({\it solid lines}) and 1.2 arcsec
({\it dotted lines}).}
\label{fig:fract}
\end{figure}

The spectra of the five templates, labeled by the numbers 0 to 4, are
shown in Figure~\ref{fig:seds} in the wavelength region of interest for
this work. Template 0 is the spectrum of
the oldest population and was build using the SPEED models, which have a
different physical treatment for the giant and horizontal branches than
the CB08 models.
The average stellar population age is
gradually reduced for the other templates, which use the CB08 models.
We explain in the next
section how we simulate the galaxy photometry using these spectral
templates.
We use the radius-luminosity relation measured in SDSS~\cite{Shen,Brown},
\begin{equation}
\log(r_e) = 1.0 - 0.26 (M_B - 5 \log h + 21.81 ) ~,
\end{equation}
where the effective radius $r_e$ is expressed in $h^{-1}\, {\rm kpc}$
and $M_B$ is the B-band absolute magnitude. We include passive
luminosity evolution (with fixed radius) as described by a \cite{BC03}
model where star formation starts at $z=4$ and decays
with a timescale $\tau = 0.6$ Gyr. This model provides a good match
to the color evolution of LRG, as reported by \cite{Brown}.
A fixed aperture of 2 ${\rm arcsec}^2$ will be assumed to simulate
the galaxy photometry. This represents a reasonable compromise between
minimizing the amount of galaxy light lost outside the aperture, and
the noise contributed by the sky. Photometric redshifts are also most
accurate when measured on fixed apertures, rather than variable
apertures that may be adjusted to the observed galaxy profile. We show in Figure~\ref{fig:fract}
the fraction of galaxy light included within the aperture as a function of luminosity for four different
values of redshift and two different seeings.

\section{Survey Model}

Throughout this paper we consider a fiducial narrow-band photometric
survey as an example of the accuracy that can be achieved to measure
a large number of LRG photometric redshifts, for the purpose of
detecting BAO. We focus on narrow-band photometry in a wavelength range that is most
useful for obtaining LRG redshifts over the range $z=0.5$ to $z=1.0$,
using the $4000\, {\rm \AA}$ (or H$\alpha$) break, a blend of H and
Ca lines. We choose a set of $N_f$ filters covering the fixed
wavelength range from $\lambda_1 = 5300\, {\rm \AA}$ to 
$\lambda_n = 8300\, {\rm \AA}$, dividing
this range into $N_f$ intervals of equal wavelength width. We will 
consider different values $N_f$ to optimize this number for photometric redshifts for a given survey and telescope setup. 
The shape of the filter window functions is assumed to be a top-hat
of width $\Delta\lambda = (\lambda_n - \lambda_1)/(N_f+1/2)$\footnote{The 1/2 is due to the presence of the wings.}, with the   
addition of lateral wings on each side where the window function varies
linearly from zero to the value in the central top-hat. The width of each of the wings is set to 1/4 of the width of the top-hat. The top-hat
parts of the window function are adjacent and non-overlapping, whereas
the wings cause the filters to have a certain degree of overlap. In a
practical application, it would probably be useful to complement this
filter system with wider filters around the wavelength range considered
here: this would eliminate some of the photo-z catastrophic failures
(outliers in the photometric redshift error distribution). However, the
addition of wide filters would not improve the accuracy of the good
redshifts because they cannot contribute to a better resolution.

\begin{figure}
\includegraphics[width=\textwidth, angle=0]{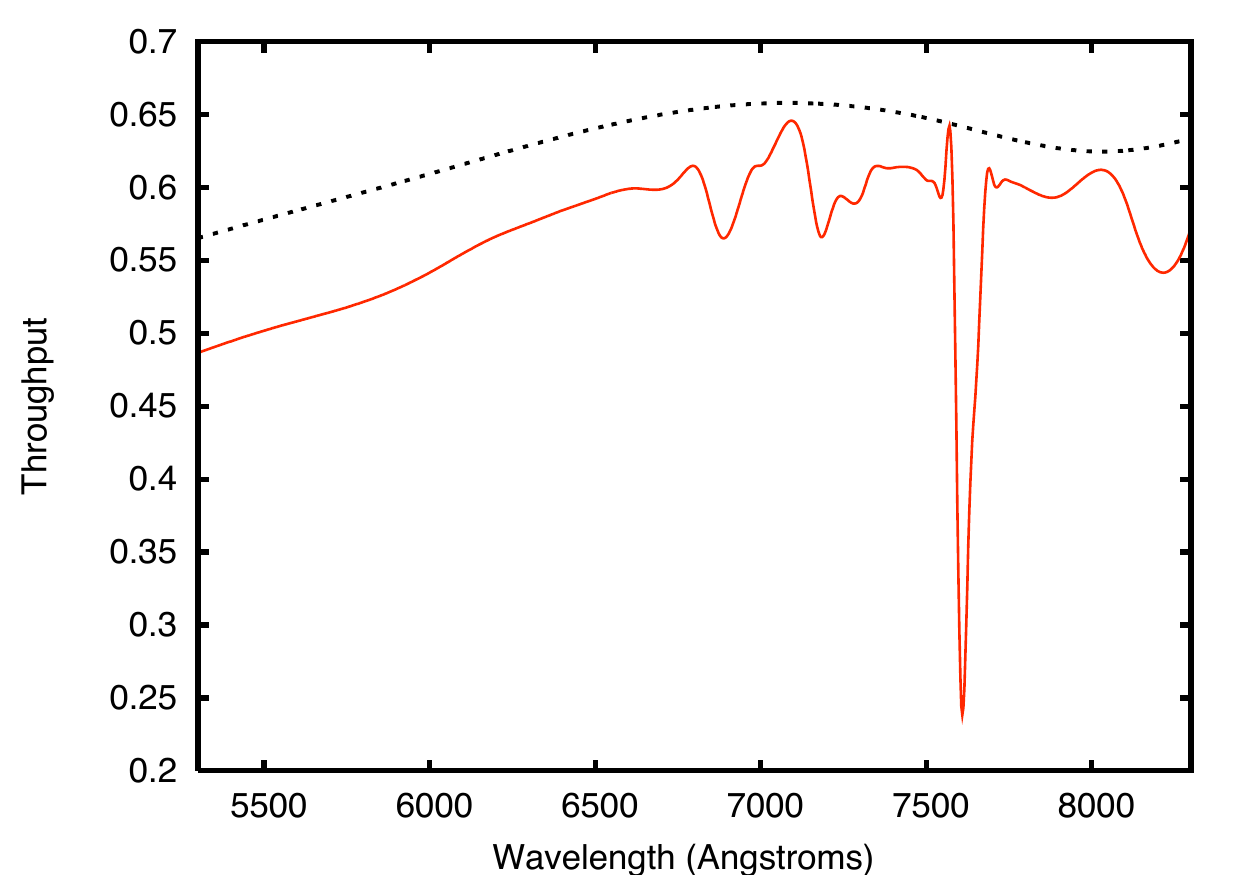}
\caption{Overall throughput ({\it solid}) and throughput without including 
atmospheric absorption ({\it dotted}) used in this work.}
\label{fig:throughput}
\end{figure}

Figure~\ref{fig:throughput} shows the overall throughput as a function
of wavelength.
We have included atmospheric absorption assuming an average of 1.2
atmospheric columns. We also include two mirror reflections, filter
transmission and CCD efficiency from \cite{royclarke, barr, ccd}.
\begin{figure}
\includegraphics[width=\textwidth, angle=0]{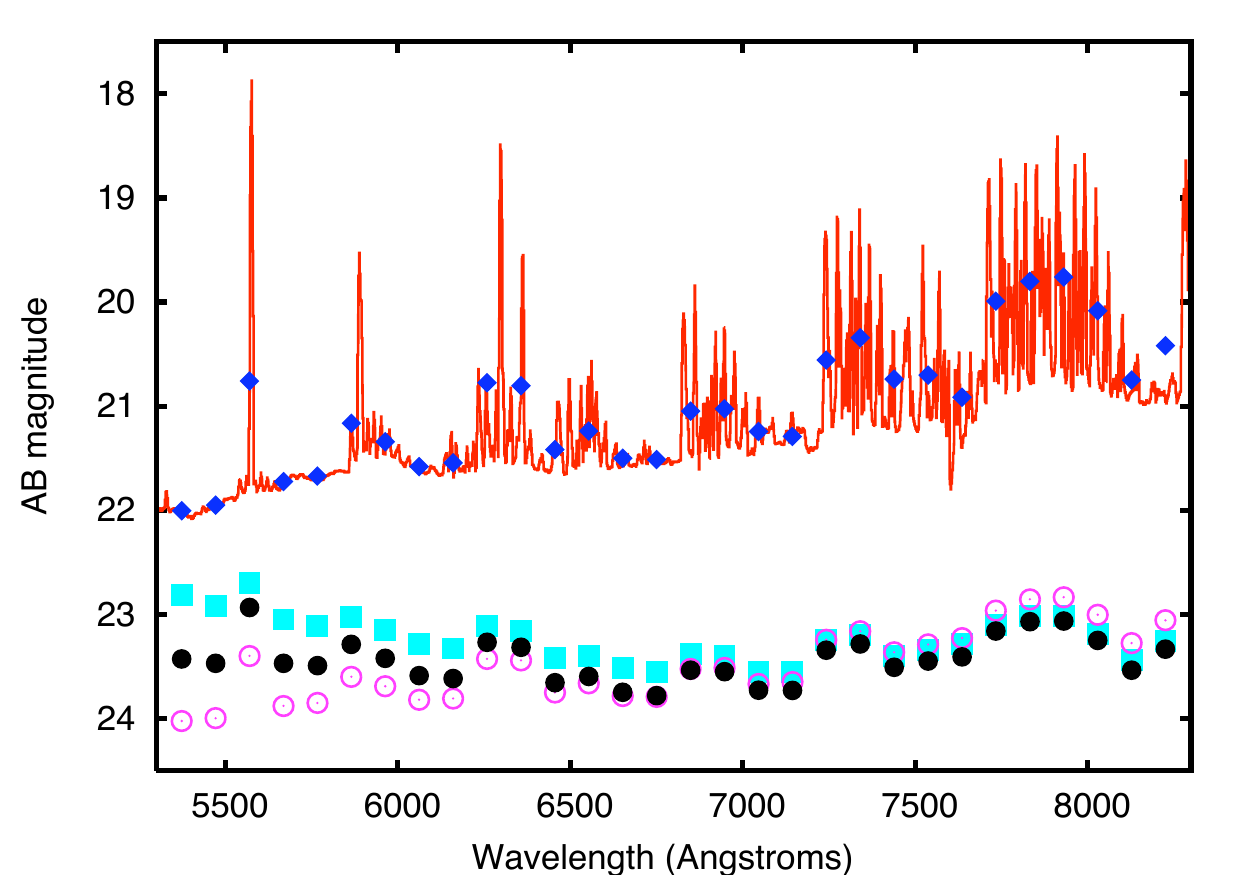}
\caption{ {\it Solid line}: Paranal sky spectrum from \cite{patat}.
{\it Diamonds}: Sky surface brightness in AB mag/arcsec${}^2$ in 
30 equally-spaced narrow-band filters. 
{\it Empty circles}: $5\sigma$ sky noise for an aperture of 2 arcsec${}^2$
in 30 narrow-band filters using a constant exposure time in our fiducial
survey. 
{\it Full circles}: $5\sigma$ sky noise for an aperture of 2 arcsec${}^2$
in 30 narrow-band filters using a linearly growing exposure time, with 
five times longer exposure in the reddest filter than in the bluest one. 
{\it Squares}: $5\sigma$ magitude limits, with the same exposure times
as for the full circles.} 
\label{fig:sky}
\end{figure}
We use the sky brightness of Patat (2008, private communication;
\cite{patat}), measured for the Paranal Observatory, to simulate the
photometric noise in each band. This
is adequate for a dark site in the absence of any artificial light, and
a mean airglow intensity over the solar cycle.
Figure~\ref{fig:sky} shows the spectrum of the sky in the AB
magnitude system, and the average sky brightness in each one of the
filters for the case $N_f=30$ ({\it blue diamonds}).
The squares and circles indicate the noise levels in our fiducial survey
and are discussed below.

Once the shape of the filters, the overall throughput, sky brightness,
source flux and radius, and seeing have been fixed, the photometric
precision reached by a survey for the flux in each filter is
proportional to $[Et/(N_f\Omega)]^{1/2}$, where $E$ is the etendue (the
product of the effective telescope aperture times the field of view),
$t$ is the total survey observing time, and $\Omega$ is the total solid
angle covered by the survey. We shall assume for our fiducial
survey a product $Et = 1.5\times 10^5 \,{\rm m}^2 \,{\rm deg}^2
\,{\rm hr}$, corresponding for example to a characteristic case of a
dedicated telescope with etendue $E=30$ (e.g., a $3$ m telescope with a
$5$ square degrees field of view) that can obtain good photometric data
for $1000$ effective hours of observation per year, over a period of $5$
years. We assume $\Omega = 8000$ square degrees, close to the maximum
area observable from a non-equatorial site, since, in general, the best
strategy for measuring BAO is to cover the widest possible area. We
also assume an optimistic value for the seeing of $0.8''$, corresponding
to the best observatories in the world. Figure \ref{fig:fract} can be
used to see how the
signal-to-noise is degraded for larger seeing: the required value of
$Et$ to reach a fixed signal-to-noise scales as the inverse square of
the fraction of light within the aperture. For example, a change of the
seeing from $0.8''$ to $1.0''$ corresponds to a $\sim$ 10\% reduction
in the fraction of light within the aperture, which must be compensated
by an increase of 20\% in $Et$.

The number of useful observing hours per year obviously depends on the
quality of the observing site. The total number of night hours per year
is $\sim 3500$; the number of observed hours is reduced owing to cloudy
or non-photometric (i.e., affected by cirri) nights, and Moon time.
Ideally, the presence of increased sky brightness (either artificial or
lunar) simply means that the effective observed time is reduced as the
inverse of the sky brightness, because the photometric noise is
inversely proportional to the square root of the ratio of the sky
brightness over the exposure time, when the noise is dominated by the
sky. For example, an increase by half a
magnitude of the sky brightness needs to be compensated by an increase
of $Et$ by a factor $10^{0.2}\simeq 1.6$. This implies an effective
reduction of the observed time due to the Moon of $\sim 25\%$, averaged
over the Moon cycle, in the wavelength range of interest.
Our assumption of 1000 effective hours of observation per year therefore
assumes that $\sim 40\%$ of the night time is clear and with acceptable
photometric conditions.
The need to co-add images obtained under different seeing, sky
brightness and transparency conditions also needs to be taken into
account as an added difficulty. In order to do faint galaxy photometry,
one often needs to discard the images with the worst seeing and convolve
the rest to a common maximum seeing, in order to avoid systematic
photometric errors that depend on galaxy morphology and may tend to
introduce artificial correlations in the photometric redshift errors.
The assumptions we make here should therefore be
regarded as optimistic and corresponding to an excellent observing site.

For our chosen value of $Et=1.5\times 10^5$, and seeing of $0.8''$,
Figure~\ref{fig:sky} shows the $5-\sigma$ sky noise in AB magnitudes
within a fixed aperture of 2 ${\rm arcsec}^2$. The open circles show the
sky noise when the total exposure time is divided equally among $N_f=30$
filters, and the filled circles assume the exposure time is a linearly
increasing function of wavelength to reduce the noise in the reddest
filters and achieve a more uniform LRG redshift accuracy over the range
$0.5 < z < 0.9$. For the latter case, we set the exposure time for the
reddest filter at 5 times that for the bluest one. We have found these
distribution of exposure times to be a reasonable compromise to achieve
the best redshift accuracies for the largest number of galaxies. We
shall use these variable exposure times for our fiducial survey
throughout the paper, with 5-$\sigma$ sky noise indicated by the filled
circles. To compute the total noise in our simulated photometry, we add
quadratically the sky, source and read-out noise. The read-out noise in
a photometric mesurement depends on the number of exposures, the pixel
size and the quality of the CCDs. We model the read-out noise as 7
electrons per pixel at each exposure, with a pixel size of 0.4 arcsec.
We assume that three exposures are obtained for each field and filter
(in practice the number of exposures could not be reduced below 3 in
most filters if the telescope operates in passive drift-scanning mode,
the best strategy to minimize photometric calibration errors). The
magnitude of a source detected at a signal-to-noise of 5 under these
assumptions is shown as the filled squares in Figure \ref{fig:sky}.
The squares approach the filled circles (the 5-$\sigma$ sky noise)
as the read-out and source noise become small compared to the sky noise.

\begin{figure}
\includegraphics[width=\textwidth, angle=0]{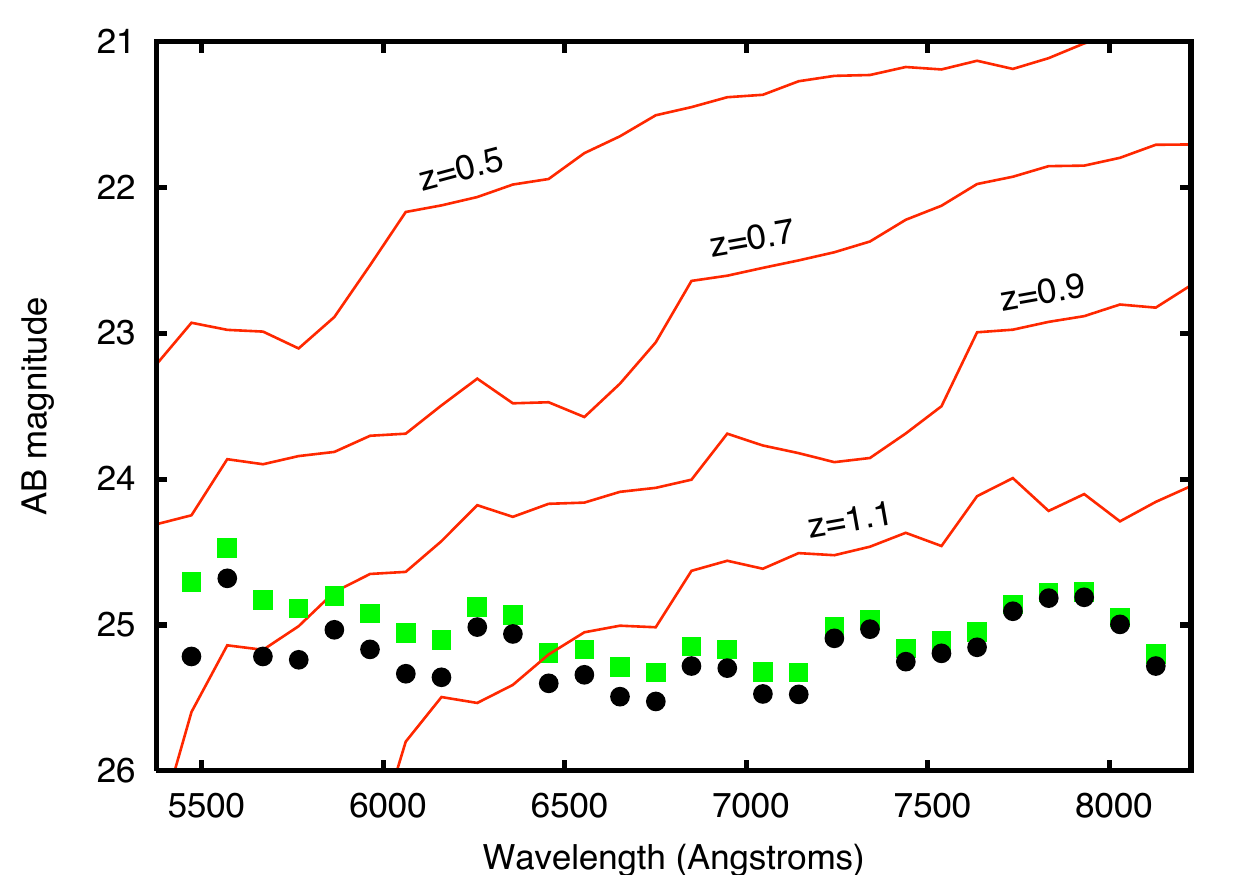}
\caption{Simulated photometry of $L_{*}$ galaxies using our template 1
at $z = 0.5, 0.7, 0.9, 1.1$ ({\it solid lines}), and the $1\sigma$ sky
noise ({\it circles}) and magnitude limits ({\it squares}).}
\label{fig:skylimit}
\end{figure}

The 1-$\sigma$ sky noise is shown in Figure \ref{fig:skylimit} as
filled circles. The
curves show the AB magnitudes of an $L_*$ galaxy in each of the $N_f=30$
filters, at four different redshifts. Note that the true S/N measured
for an $L_*$ galaxy is slightly decreased compared to that shown in this
Figure because of the source photons contribution to the total noise;
this is only a $\sim$ 10\% effect for an $L_*$ galaxy at $z=0.5$ on the
first filter to the red of the $H\alpha$ break.

To model the accuracy of photometric redshifts, we proceed as
follows in the next section: first, we simulate the $N_f$-filter
photometry of a galaxy that has the spectrum of one of the five
templates we use. We add noise for a survey
with the characteristics  described above, and fit the resulting
magnitudes with  the same template.  This will tell us the
minimum photometric redshift error that would be obtained in the ideal
case of a galaxy for which we know exactly the spectral shape, and we
fit only one parameter, the redshift, to the measured photometry. We refer to 
this as ``single template case". We shall examine the dependence of the results 
on the template used, and comment on the effects of galaxy evolution.
We shall then consider the case where a linear combination of the five
templates is used to fit to the photometry simulated with one template
(5-templates case) and the case where the full set of 5 templates is
used to simulate and fit the photometry (5-to-5 templates case).
This will show how the redshift error increases when galaxies are
allowed to have variable spectra with five parameters (the redshift and
four relative template amplitudes), but the spectra are still assumed to
be precisely modelled by the templates used in the fit. In these two
cases, the effect of changing the number of filters will be explored.

Finally, we shall use observed spectra of LRG from the SDSS public
data to generate a photometric ``mock" catalog, and fit their photo-z's
using our 5 templates. The resulting photometric errors should be more
realistically expected for real galaxies, which are not perfectly
matched by a limited number of templates. We note that the 5 templates
we use have been optimized to fit the spectra of a wide class of
observed galaxies, and therefore they should be an ideal set for
obtaining the best possible photometric redshifts.

\section{Photometric Redshift Accuracy}

\subsection{Optimal number of filters: analytic considerations}

Before presenting our numerical results on the simulated photometry
described in the previous sections, it will be useful to derive some
analytical results on the photo-z errors and the optimization of the
number of filters under some simplifying approximations, using simple
examples of galaxy spectra. We consider here the ideal case where a
galaxy has exactly the same spectrum as the template used to fit the
measured photometry, and approximate the filter windows as a top-hat.

If a fixed total exposure time is to be divided among all the filters,
then the signal-to-noise in each individual filter, $F/\sigma_F$ (where
$F$ denotes flux and $\sigma_F$ its statistical error), is inversely
proportional to the number of filters $N_f$: the exposure time in each
filter is $t_f \propto N_f^{-1}$, and the wavelength width of each
filter is $\Delta \lambda \propto N_f^{-1}$. Hence, the number of
photons detected in each filter is
$N_{ph} \propto t_f \Delta\lambda \propto N_f^{-2}$, and
$F/\sigma_F \propto N_{ph}^{1/2} \propto N_f^{-1}$. As we increase the
number of filters, the resolution increases at the expense of the
achieved signal-to-noise.

To model the use of the H$\alpha$ break in LRG, we first consider a
galaxy spectrum
that has a break at wavelength $\lambda_0$, where the flux per unit
wavelength is $F_{\lambda} = F_0$ at $\lambda > \lambda_0$ and
$F_{\lambda} = (1-B) F_0$ at $\lambda < \lambda_0$. The filter that
includes the break wavelength $\lambda_0$ has a flux
\begin{equation}
F = F_0 (1-Bx) ~,
\end{equation}
where $x=(\lambda_i - \lambda_0)/\Delta\lambda$, and $\lambda_i$ is
the wavelength of the right edge of the filter.
From the measured value of $F$ (and assuming that we know exactly the values of $F_0$ and $B$
from the measurements in the other filters), the wavelength $\lambda_0$
can be measured to an accuracy
\begin{equation}
\sigma_{\lambda} = \Delta\lambda\, {\sigma_F\over F_0 B} ~.
\end{equation}
Since $\sigma_F/F_0 \propto N_f$, and $\Delta \lambda \propto N_f^{-1}$,
we conclude that the error to which the wavelength $\lambda_0$ is
measured is independent of the number of filters, and results in a
photometric redshift accuracy $\sigma_z$ given by
\begin{equation}
{ \sigma_z\over 1+z} \equiv \sigma_{lz} =
{\sigma_{\lambda}\over \lambda} = {\Delta\lambda \over \lambda}\,
{\sigma_F \over F_0 B} ~.
\label{errpz}
\end{equation}
In other words,
unresolved breaks in the spectra of galaxies yield a redshift
measurement that does not improve as the number of filters is increased.
This is correct only in the idealized situation where the amplitude of the
break is perfectly known and galaxy variability may be ignored.

For a resolved break, we assume that the flux varies linearly from $F_0$ to
$F_0(1-B)$ over a wavelength range $\lambda_0$ to $\lambda_0+ \delta\lambda$.
The flux measured in a filter centered at wavelength $\lambda_i$ and which
is fully included within the interval of the break of width
$\delta\lambda$ is $F=F_0(1-Bx)$, where
$x = (\lambda_i-\lambda_0)/\delta\lambda$. Hence, the error on
$\lambda_0$ that can be deduced from the flux measured in one filter
only is
\begin{equation}
\sigma_{\lambda} = \delta\lambda\, {\sigma_F\over F_0 B} \propto N_f ~.
\end{equation}
Since the number of filters contained in the wavelength range of the
resolved break, $\delta\lambda$, is proportional to $N_f$, and the
combined error from the measurement in all the filters is reduced as
$N_f^{-1/2}$, we find that the set of all photometric measurements will
yield an error $\sigma_{\lambda} \propto N_f^{1/2}$.
Therefore, the filters should
not be any narrower than the intrinsic width of any breaks that
substantially contribute to the photometric redshift measurement.

On the other hand, if the spectrum contains features similar to
an emission or absorption line, then it is advantageous to
increase the number of filters. For a line feature with an equivalent
width $W_\lambda$ which is entirely contained in one filter, the
measured flux is $F=F_0(1+W_\lambda/\Delta\lambda)$, independently of
the position of the line within a top-hat filter. The difference
$F-F_0 = F_0 W_{\lambda}/\Delta\lambda \propto N_f$ can be measured to
an accuracy $\sigma_F \propto N_f$, so the position of the feature is
measured to increasing accuracy as the number of filters is increased,
up to the point where the feature is resolved.

In practice, the spectra of LRG contain several features that may
be approximately modelled as breaks and/or line features of different
wavelength widths. In the case of a single template, the optimal number
of filters results from the contribution of all the features to the
determination of the photometric redshift. At the same time, the
spectra of real galaxies are, of course, not described by a single
template and depend on a variety
of parameters. The need to fit for these spectral variations
in addition to the redshift favors an increased number of filters when
the signal-to-noise that can be reached is high enough. This is seen in
more detail in the following section; the analytic considerations
discussed above will be helpful to interpret the numerical results for
fits to simulated photometry of LRG.

\subsection{Single template galaxies}

We now present the numerical results of recovering photometric
redshifts from the simulated photometry of LRG spectra, for the case of
our fiducial survey as described in \S 4. 
 The distribution of photometric redshift errors is generally not
Gaussian because of the presence of outliers, or catastrophic failures,
which naturally increase as the signal-to-noise drops. To avoid having
to deal with identification of outliers, we report the photo-z error
$\sigma_z$ in all of our figures, unless otherwise stated, as the
interval such that 68\% of the simulated galaxies have a smaller
difference between their true redshift and fitted redshift. As long as
the fraction of outliers is small, the 68 percentile value should
approximately correspond to the rms redshift error after the outliers
have been successfully removed. We shall not discuss here the methods
for removing outliers in photometric redshifts; these would depend on
how our assumed narrow-band filter system is complemented with wider
filters on a broader range of wavelengths. In practice, any method for
fitting photometric redshifts will fail to detect some outliers
and will classify some good redshifts as outliers. Our assumption that
outliers are removed perfectly is the most optimistic possible one.

\begin{figure}
\includegraphics[width=\columnwidth, angle=0]{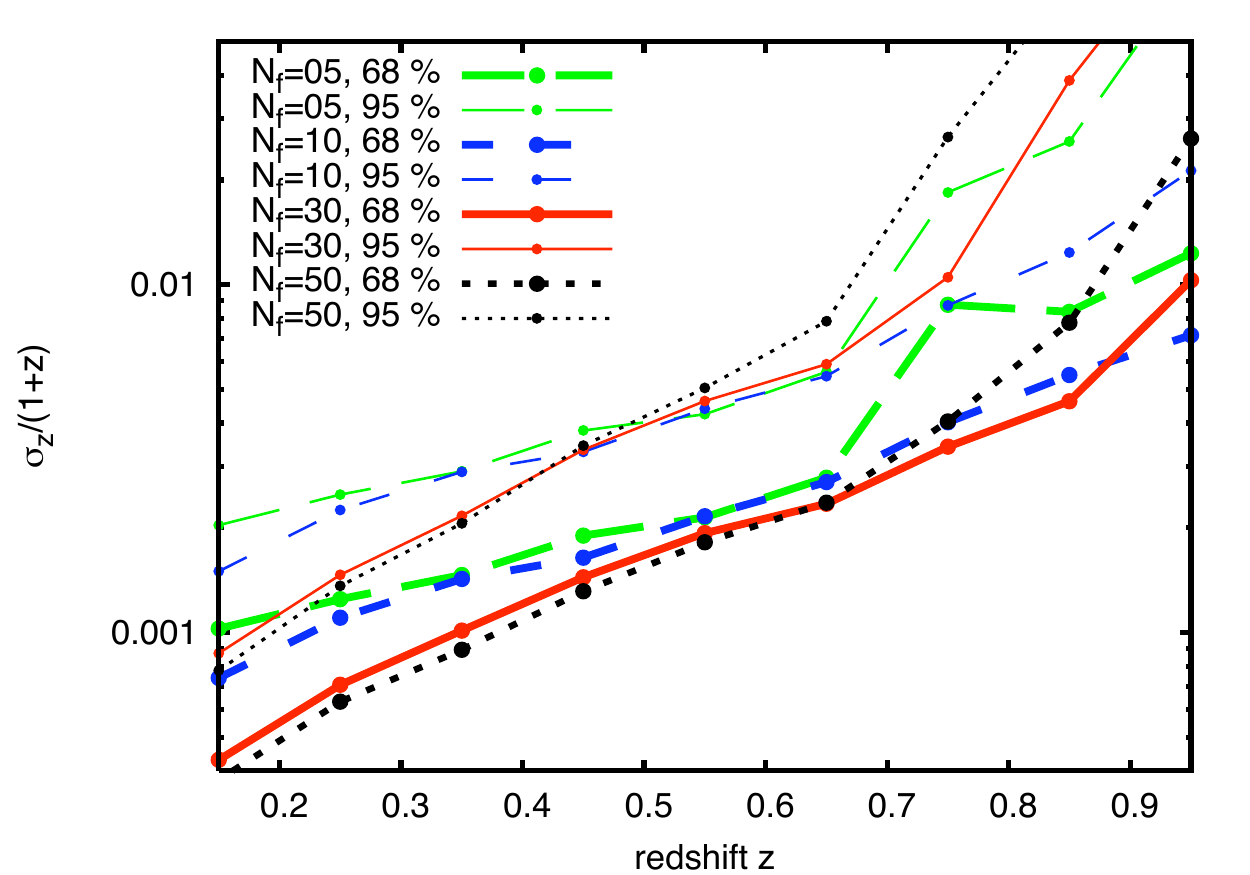}
\includegraphics[width=\columnwidth, angle=0]{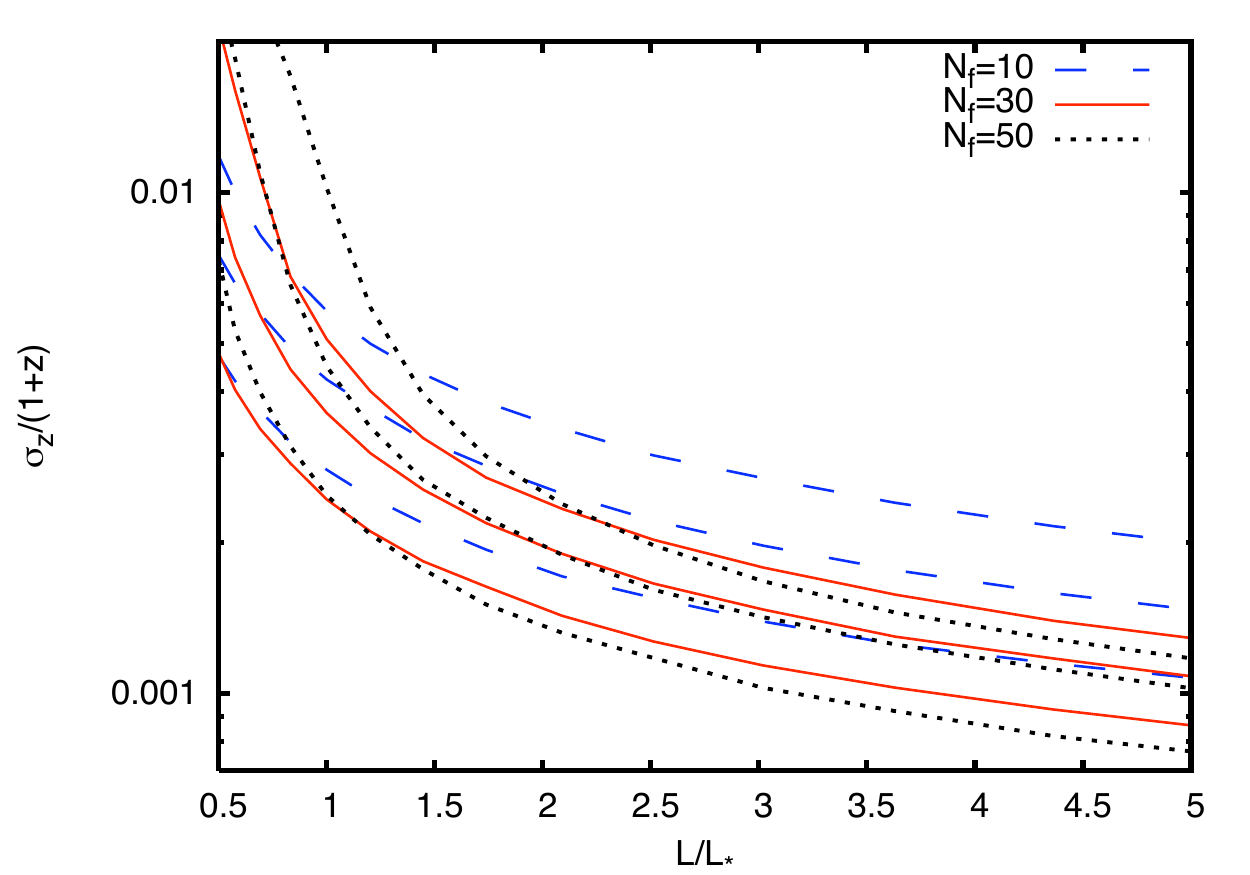}
\caption{{\it Top}: Photo-z error as a 
function of redshift of an $L_*$ galaxy in our fiducial survey,
varying the number of filters as indicated.
The 68 or 95 percentile error is computed in each redshift bin of
width $\Delta z = 0.002$, and then averaged in quadrature over
$\Delta z = 0.1$.
{\it Bottom}: Photo-z error (68 percentile, computed as in the top
panel) as a function of galaxy luminosity,
for redshift intervals centered at $z=0.65$, 0.75 and 0.85 (errors
increase with redshift), and for $N_f=10$, 30 and 50.
Template 1 is used in all cases for generating and fitting the galaxy
photometry.}
\label{fig:f5}
\end{figure}

 We start with the simplest case where the same template used to
generate the galaxy photometry is then used to fit the photometric
redshift. This corresponds to the ideal case where the galaxy spectral
shape is precisely known, and only the redshift and luminosity need to
be fitted. We use template 1 for this purpose in this section. The
relative redshift error $\sigma_{lz} = \sigma_z/(1+z)$ is shown in
Figure~\ref{fig:f5} (top panel) as a function of redshift, for galaxies
of luminosity $L_*$ and varying the number of filters. For this Figure,
we show both the 68 percentile error ({\it thick lines}) and the 95
percentile error ({\it thin lines}). The thin and thick lines should be
separated by a factor 2 for a Gaussian distribution of errors, and by an
increasing factor as the fraction of outliers increases. The redshift
errors are calculated at 450 values of the redshift, from $z=0.1$ to
$z=1$ in increments of $\delta z = 0.002$. For each value of the
redshift, the error is computed from a total of 100 photometry
simulations. Therefore, each curve in Figure \ref{fig:f5} (top panel) is
based on a total of 45000 simulations of galaxy photometry.
The results are shown as filled dots, after averaging in quadrature over redshift
intervals of width $\Delta z = 0.1$ (lines between dots are plotted for
guidance only). The bottom panel of Figure \ref{fig:f5} shows the
redshift error (68 percentile) as a function of galaxy luminosity
for redshifts 0.65, 0.75 and 0.85 (with errors increasing with redshift),
again after averaging over an interval $\Delta z = 0.1$, and for
$N_f=10$, 30 and 50.

 We see from this Figure that for galaxies of luminosity $L_*$, the
redshift errors are practically independent of the number of filters in
our fiducial survey. At $z>0.4$, there is only a small improvement of
$\sigma_z$ when $N_f$ is increased from 10 to 30, and even this small
improvement is offset by an increasing fraction of outliers as $N_f$ is
increased. This implies that it would not be worth to use more than
$\sim$ 10 filters to measure photometric redshifts for galaxies of
luminosities near $L_*$ that are well fitted by a single template.

 The target accuracy of $0.003(1+z)$ to measure radial BAO is achieved
up to $z \sim 0.7$ for $L=L_*$. The bottom panel shows that at $z=0.9$
the same target can be achieved for $L > 1.6 L_*$. For galaxies of high
luminosity, increasing $N_f$ from 10 to 30 becomes a stronger advantage,
but gains for $N_f > 30$ are not substantial, especially because the
outliers increase with $N_f$. The approximately constant redshift
accuracy with the number of filters is roughly a consequence of the
analytic arguments discussed in \S 5.1 for the case when the main
spectral feature is an unresolved break. Note that there are also
numerous features in the spectrum of template 1 (Figure 2) that can be
modelled as absorption lines of width $\sim \lambda/30$ (particularly
at wavelengths just blueward of the H$\alpha$ break), which are the
principal reason for the improvement of redshift errors from $N_f=10$
to $N_f=30$ for high signal-to-noise.

\begin{figure}
\includegraphics[width=\columnwidth, angle=0]{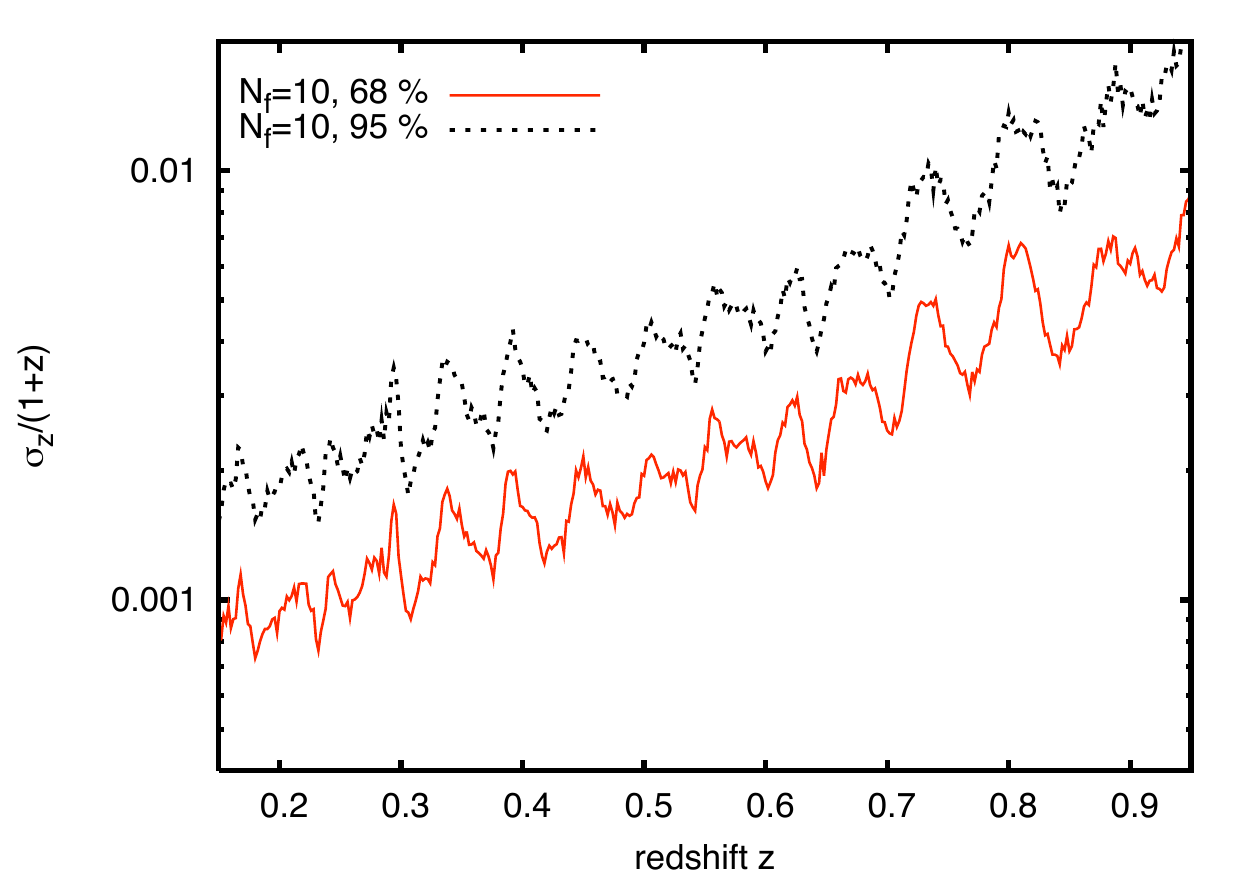}
\includegraphics[width=\columnwidth, angle=0]{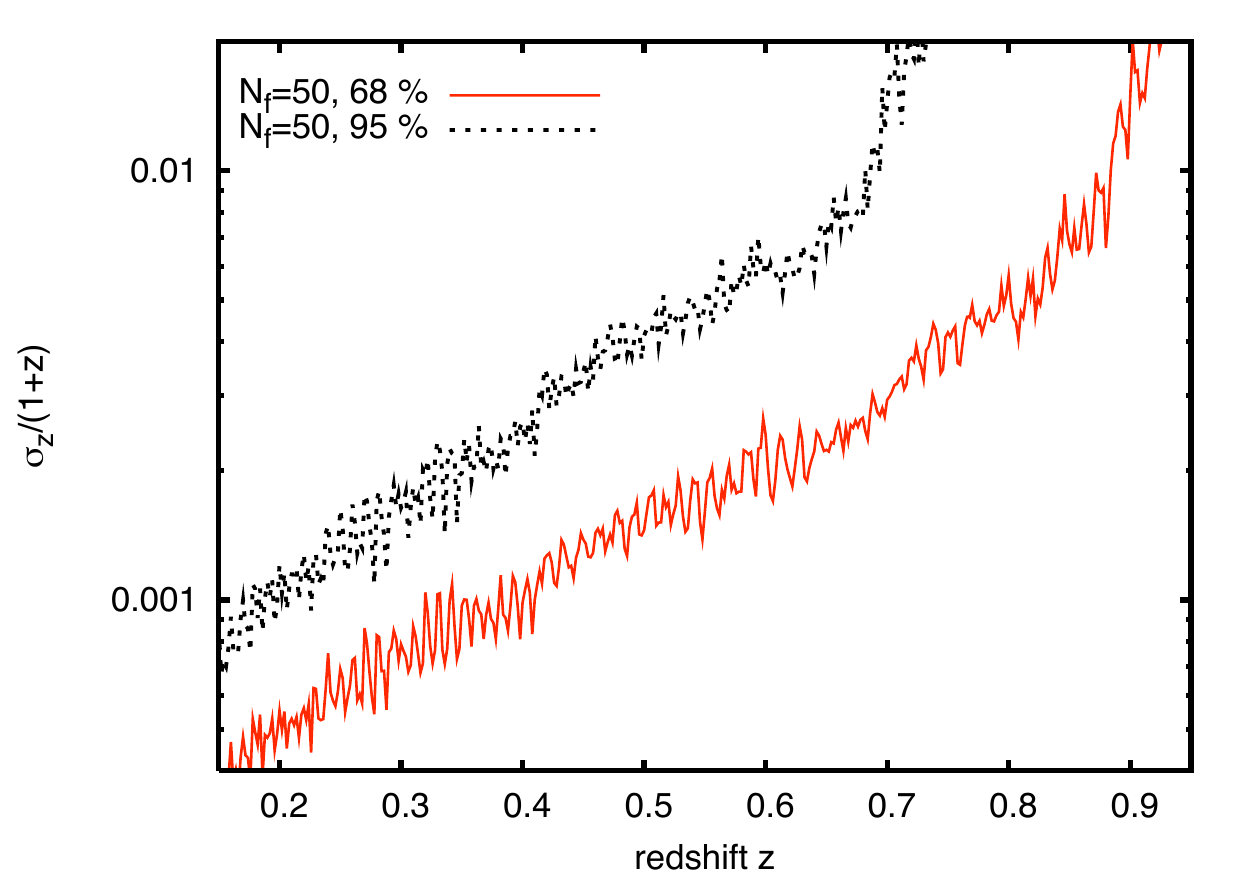}
\caption{Photo-z error as a function of redshift without smoothing for
10 filters ({\it top}) and 50 filters ({\it bottom}). The
oscillations have a periodicity determined by the filter width.}
\label{fig:f7}
\end{figure}

 So far, the results have been shown averaged over redshift bins of
$\Delta z = 0.1$. The unbinned results are shown in Figure~\ref{fig:f7},
for the cases $N_f=10$ and $N_f=50$. There are periodic
oscillations in the redshift error, with a spacing $\delta z$ that
corresponds to the separation between filters. These oscillations are
due to different features in the spectrum (mostly the H$\alpha$ break)
crossing the center and the edge of filters as the redshift varies. The
redshift error is best (smallest) when the H$\alpha$ break is placed
between two filters, and it is worst when it is placed at the center of
one filter. These oscillations should naturally depend on the shape of
the filter windows, as well as the form of the H$\alpha$ break, and
therefore the type of LRG galaxy. One should note that the scale of
the oscillation corresponds to the BAO scale for $N_f \simeq 30$. The
presence of these artificial oscillations in the redshift error, which
could vary their amplitude over the survey area owing to variations of
the seeing or other observing conditions that may systematically affect
the photometry, needs to be properly calibrated and corrected for in
order to avoid introducing perturbations in the BAO peak of the galaxy
correlation function.

\subsection{Using multiple templates}

 In reality LRG are not a perfectly uniform population described by a
single spectrum. We now see the impact on the redshift errors when
variations in the galaxy spectra are allowed for. We use the five
templates described in \S 3, which have been optimized to provide a
good representation of observed galaxy spectra \cite{Niemack,Panter07}.

\begin{figure}
\includegraphics[width=\columnwidth, angle=0]{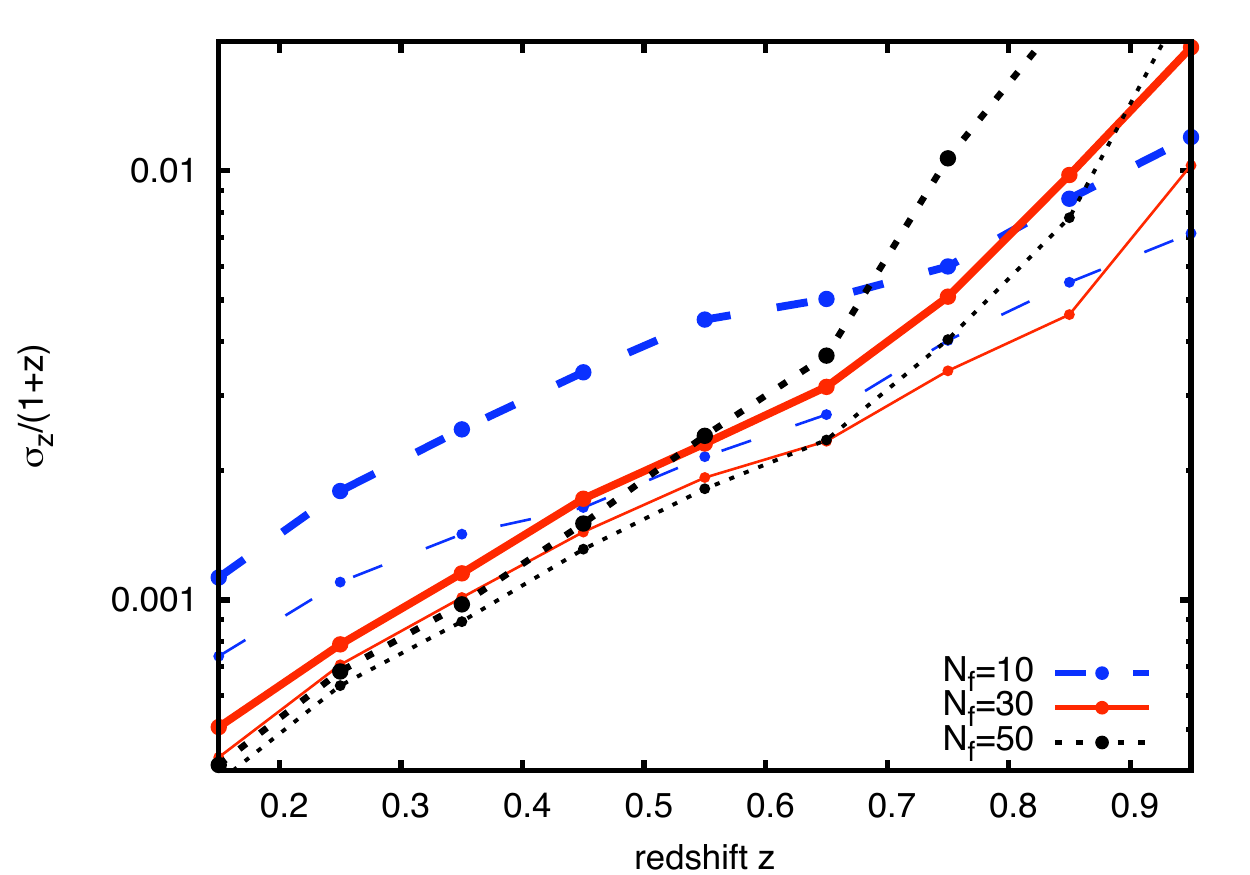}
\caption{Impact of using multiple templates on redshift errors.
{\it Thin lines}: The mock $L_*$ galaxy photometry is generated using
template 1 and fitted with the same template. {\it Thick lines:} The
same mock catalogs generated with template 1 are fitted with all five
templates. The 68 percentile errors are computed as in
Figure \ref{fig:f5}. The number of filters is varied as indicated.}
\label{fig:f10}
\end{figure}

 Figure \ref{fig:f10} shows the photo-z error for spectra generated with
template 1 and $L=L_*$. The thin lines are the same as those shown in
Figure \ref{fig:f5} (top panel), fitted with the same template 1. The
thick lines are the result obtained when any linear combination of the
five templates is allowed in the fit. As expected, the errors increase
when we allow for spectral variability: the maximum photo-z accuracy
for an ideal galaxy having exactly the same spectrum as template 1 is
obtained when we know that the galaxy has the spectrum of template 1.
When we do not know this, and spectral variability needs to be allowed
for, the photo-z error is larger. When fitting with our five templates,
the target accuracy of $\sigma_z = 0.003 (1+z)$ is achieved only for
$z< 0.63$ at $L=L_*$ for $N_f=30$ (which is roughly the ideal number of
filters). Results for $L > L_*$ will be shown below in Figure
\ref{fig:flum}.

 The increase of the redshift error resulting from fitting with five
templates is substantially worse for $N_f=10$. When LRG are not assumed
to be a homogeneous population, using a larger number of filters becomes
more advantageous because of the need to distinguish the type of galaxy
spectrum when fitting the redshift. However, as found in the previous
section, increasing further the number of filters to $N_f=50$ is
generally not worthy, except at rather high signal-to-noise, where the
number density of sources is small in our fiducial survey.

 The results in Figure \ref{fig:f10} are a lower limit to the
increase of redshift errors caused by galaxy variability, because real galaxies
are not perfectly matched by the set of five templates we use. 
We discuss this further in \S 5.5.

\subsection{Variation of redshift error with templates and effects of
galaxy evolution}

\begin{figure}
\includegraphics[width=\columnwidth, angle=0]{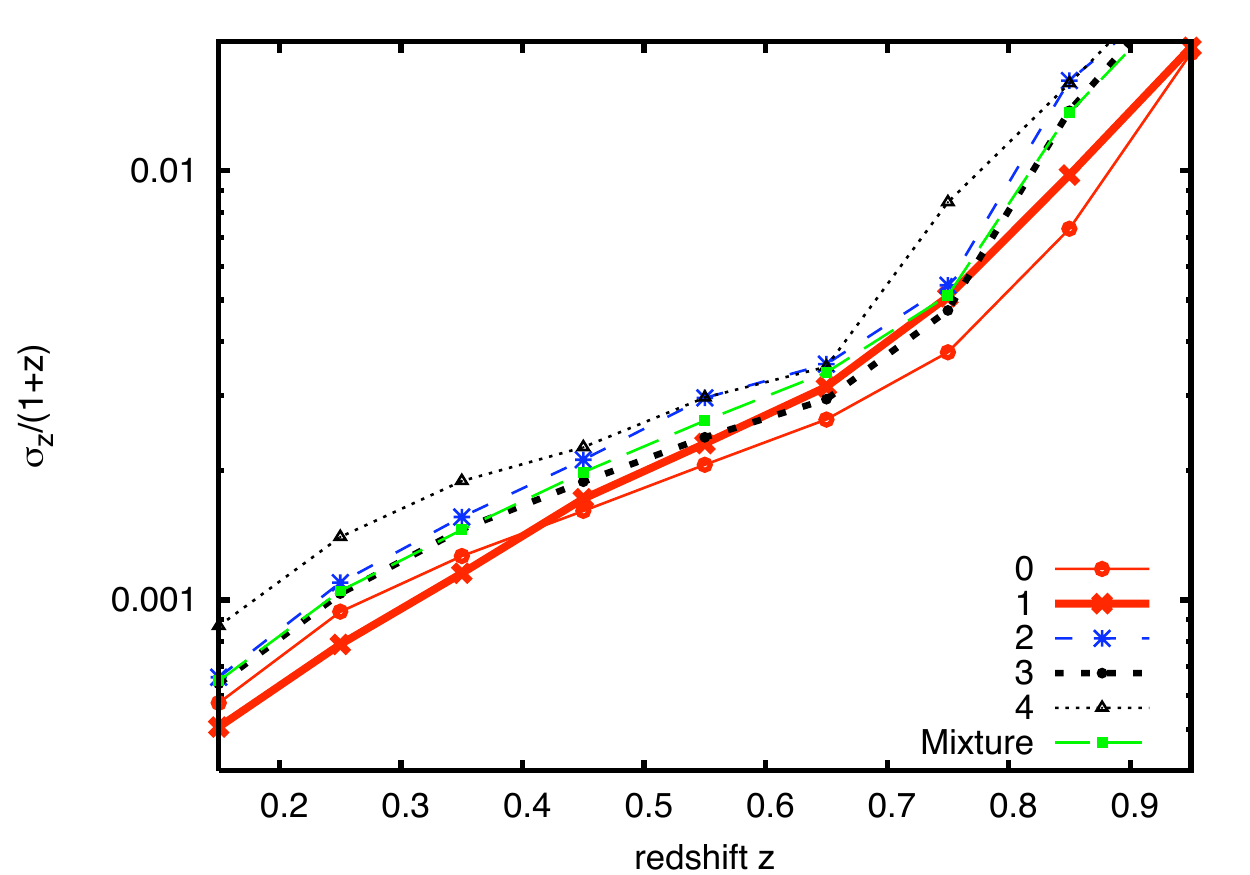}
\includegraphics[width=\columnwidth, angle=0]{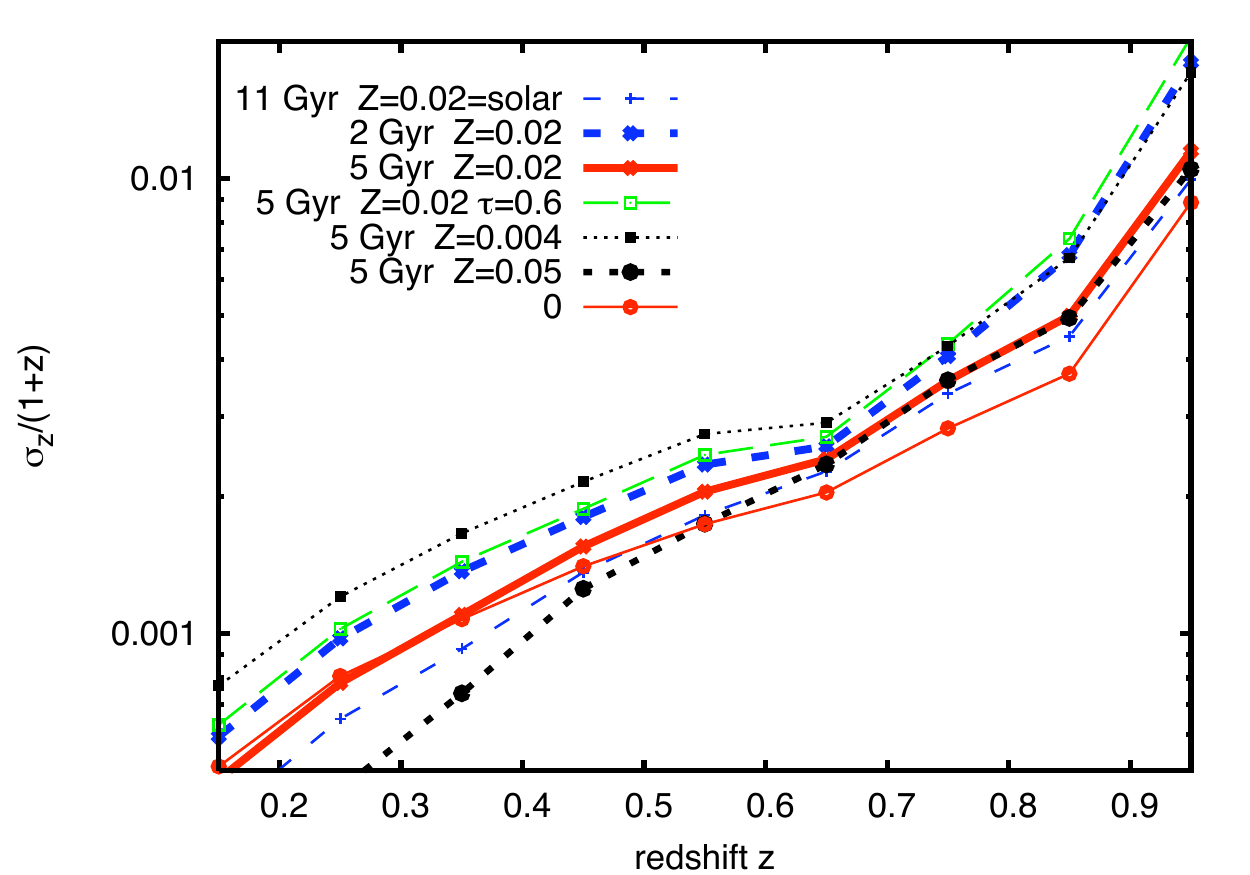}
\caption{ {\it Top panel}:
Mock catalogs of $L_*$ galaxies are generated with one template for
each line (0-4) and fitted with the five templates. We also show the
case where the mock catalog is generated by random linear combinations
of the five templates and fitted with the five templates.
{\it Bottom panel}: Photo-z accuracy for spectra of single starbursts of
fixed age and metallicity and no dust, compared with template 0. The
accuracy improves with increasing age and metallicity, and worsens with
even trace amounts of star formation, but depends also on details of the
modeling of giant and horizontal branch stars.}
\label{fig:f11}
\end{figure}

 Figure \ref{fig:f11} ({\it top panel}) shows the photo-z errors when
the galaxy photometry is generated for each one of the five templates,
and fitted allowing a linear combination of all five templates, for
$L=L_*$ and $N_f=30$. 
There are substantial variations of
the redshift error with the template spectrum, implying that the
redshift error depends on the type of LRG galaxies selected to measure 
BAO. Over the redshift range $0.5 < z < 0.9$, which is the most promising for 
our purpose,
template 0 gives the smallest errors and
templates 2 and 4 the largest ones. The long-dashed line shows the error
when random linear combinations of the five templates are used to
generate the galaxy photometry. As expected, the error obtained is an
average of the errors from the templates 0 to 4.

 To understand the  reasons why the errors depend on the
template used, we plot in the bottom panel of Figure \ref{fig:f11} the
photo-z error for a set of templates computed for a stellar population
of fixed age (11, 5 and 2 Gyr) and metallicity ($Z_{\odot}$,
$0.2 Z_{\odot}$, or $2.5 Z_{\odot}$); a template with an exponentially
decaying star formation rate with $\tau=0.6$ Gyr at age 5 Gyr is also
included. These set of templates have been generated with the model of
Bruzual \& Charlot (2003), and are used only in this figure for 
illustrative purpose. 
Smaller errors are generally obtained
for stellar populations that are older and of higher metallicity. The
presence of even a small amount of young stars (the model shown has a
star formation rate that has decreased by a factor $e^{-5/0.6} =
2.4\times 10^{-4}$ since star formation started) increases the errors. 
There are, however, other factors affecting the photo-z errors in
addition to the age and metallicity of the stellar population. In
particular, the modeling of the giant and horizontal branch stars
affects the part of the spectrum near the H$\alpha$ break, and has a
complex effect on the redshift errors. We note that, for $z>0.5$,
template 0 has smaller errors than all other templates. The variations
between templates 0 to 4 are due to differences in the distribution of
stellar ages, amount of dust absorption, and different stellar evolution
models used. Let us recall that templates 0 to 4 are built using the
empirically derived Panter et al.\ (2007) star formation histories.
Template 0 uses the SPEED models \cite{Jimenez+04}, which have been
empirically found to provide a better fit to observations of LRG spectra
than other stellar population models.
Despite having the oldest stellar population, template 0 is bluer than
template 1 in the rest-frame wavelength range of $4000$ to $9000$ \AA\ because 
it has a less developed giant branch and has no dust extinction.
Template 1 aims at reproducing the spectrum of an S0/Sa galaxy, and has
some dust absorption.
Templates 1 to 4 are built using Charlot \& Bruzual 2008 models, which
have been empirically shown to provide a better fit to Sa-Sc galaxies.
Their larger errors compared to template 0 are due to different star
formation histories (which vary among the templates in a complex way)
and the different stellar evolution model. We note also that the
redshift error depends on the band used to normalize the templates to
the same luminosity; here we have normalized them to the same B-band
luminosity, which is used to measure the luminosity function \cite{Brown}.

The variation of the errors with the assumed stellar population in
each template is roughly associated with the decreasing amplitude of
the H$\alpha$ break for younger and less metal-rich galaxies, and with
the addition of trace amounts of star formation (see \S 5.1,
and eq.\ [\ref{errpz}]). For the same reason, changes in the modeling
of giant and horizontal branch stars that affect the change of the
H$\alpha$ break have an important impact on redshift errors.
Despite the complex dependence of the photo-z errors on several
properties of the stellar population, the expected evolution with
redshift is clear: stellar populations in LRG should be younger, star
formation rates should increase, and the LRG population should become
less homogeneous, leading to increased redshift errors. For similar
reasons, redshift errors should decrease with galaxy luminosity.
There is growing observational evidence that the most massive
galaxies contain the oldest stellar populations up to $z \sim 1-2$ and
they have less than 1\% of their present stellar mass formed at $z<1$
\cite{Cowie99,Heavens04,thomas,Panter07}.
In high-density regions (i.e., galaxy clusters), massive systems
ceased their star formation by $z \sim 3$ \cite{thomas}.
While galaxies with stellar masses $ > 5 \times 10^{11}$~M$_{\odot}$
(roughly corresponding to $2L_*$) ended their star-formation activity
by $z \sim 2$, less massive
objects were still actively forming stars at $z \sim 1$ \cite{treu}.
As a result, the redshift errors found here using SDSS spectra of LRG,
which are at $z\lesssim 0.4$ and $L\gtrsim 3 L_*$, are a lower limit to
the redshift errors that may be found at higher redshifts and lower
luminosities.

\begin{figure}
\includegraphics[width=\textwidth, angle=0]{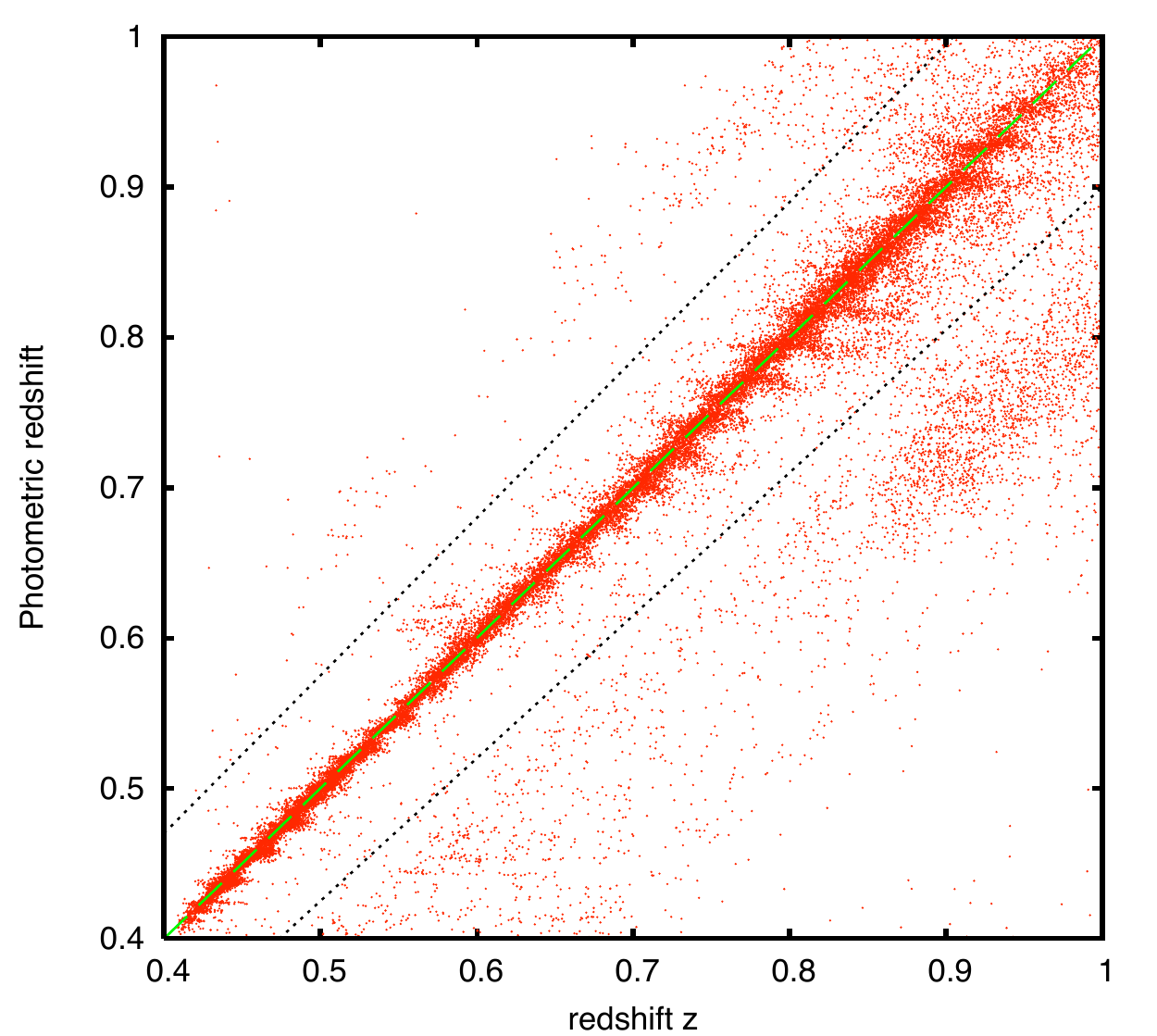}
\caption{Photometric redshift versus input redshift. Our sample is
generated by redshifting the spectra of SDSS DR6 LRG for $L=L_*$ and
our fiducial survey (see text), and fitting them with our five
templates. Dotted lines show the $\pm 0.05 (1+z)$ photometric error
range. All points outside the dotted lines are considered outliers that
could eventually be removed.}
\label{fig:f15}
\end{figure}

\begin{figure}
\includegraphics[width=\textwidth, angle=0]{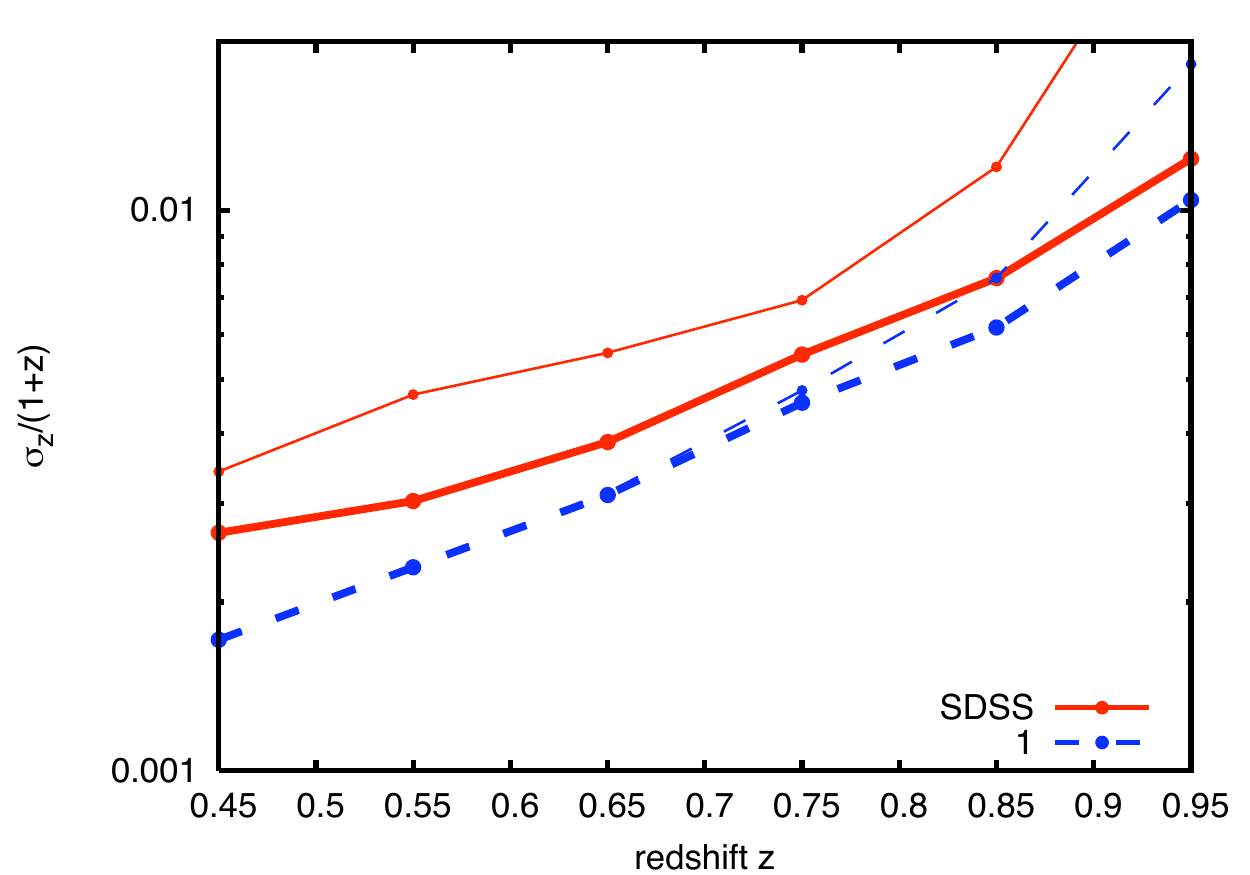}
\caption{{\it Solid lines:} Photo-z error for redshifted SDSS LRG at
$L=L_*$ in our fiducial survey, fitted with our five templates. The
thin line shows the 68\% error in each redshift interval of width
$\Delta z =0.1$ without removing any outliers, and the
thick line shows the same error when the outliers outside the range
of the dotted lines in Figure \ref{fig:f15} are removed. {\it Dashed
lines:} Same results when we use a mock catalogue generated with
template 1. The recovered photo-z errors for observed SDSS increase
relative to the case of ideal galaxies generated exactly with one of
the five templates used.}
\label{fig:f16}
\end{figure}

\subsection{Effects of variability in real LRG: SDSS Spectra}

  In practical observations, the spectra of galaxies for which
photometric redshifts are obtained are never exactly the same as the
templates used for the fits. Real galaxy spectra are affected by
the distribution in age, metallicity and dust obscuration, are not
matched well enough by our imperfect models of stellar evolution, and
may be substantially changed by very small amounts of star formation.
For example, the presence of a weak OII emission line (at a wavelength
close to that of the H$\alpha$ break) due to traces of star formation
may change the recovered photo-z. In addition, superposed light from
very faint galaxies along the line-of-sight may introduce small changes
in the spectra. Therefore, the redshift errors we found so far by
assuming that the adequate templates are perfectly known may be
severely underestimated.

  The best and most extensive available survey of LRG is the SDSS. 
In this Section we use observed spectra of LRG obtained from the
publicly available SDSS-DR6; {\tt spectro.princeton.edu} catalog.
We randomly select $50,000$ galaxies flagged as LRG for which a high
signal-to-noise spectrum is available.
  Unfortunately, most of the SDSS LRG are at redshifts $z<0.4$. To
simulate the measurement of a photometric survey at higher redshifts, we
shift all the observed SDSS spectra by a constant 
factor of $f=530/380 = 1.395$ to longer wavelength.
Hence, the true redshifts of the
galaxies, $z$, are all shifted to a redshift for the simulated galaxy
$z' = (1+z)f - 1$. With this shift factor, the minimum observed
wavelength in the SDSS spectra, at $\lambda=380$ nm, is shifted just to
the blue edge of our first filter at $530$ nm, and the redshift
range shifts to the interval $0.4 < z < 1.0$, the most interesting range
for measuring BAO with LRG photometric redshifts. We then follow the
same procedure described in \S 4: the shifted spectra
are convolved with the filter window functions, noise is added
to the fluxes at each filter, and the simulated photometry is fitted
with our five templates.

  We first examine a scatter plot of the fitted photometric redshift
versus the true redshift of the SDSS galaxies (after the constant
shifting mentioned above) for $L=L_*$ and our fiducial survey with
$N_f=30$, in Figure \ref{fig:f15}. For guidance, a dashed line for
$z=z_{phot}$ has been added. The two dotted lines indicate
$1+z_{phot} = (1 \pm 0.05) (1+z)$, and will be our threshold for
considering a photometric redshift as an outlier. The effect of the
oscillations in the redshift error distribution with a period equal to
the filter width, mentioned in \S 5.2, is also seen here. The dotted
lines clearly separate a group of outliers in the error distribution,
which arise because the H$\alpha$ break is confused with other wider
features in the spectrum when the noise is high. In the rest of this
section, the results for the 68 percentile of the redshift error
distribution will be shown for two different cases: considering the
whole redshift error distribution, and removing the outliers defined
as the points outside of the two dotted lines. The errors computed after removing
the outliers should be considered as a highly optimistic case where the
outliers are assumed to be perfectly identified with the help of
additional broad-band filters in the survey.

  The solid lines in Figure \ref{fig:f16} are the 68 percentile redshift
error as a function of redshift, for $L=L_*$ and our fiducial survey
with $N_f=30$, as in Figure \ref{fig:f15}. The thin line is the value
for the whole distribution, and the thick line is obtained after the
outliers are removed. The effect of the outliers is substantial,
increasing the redshift error by $\sim$ 40\%. For comparison, we show as
dashed lines the case where galaxies generated with template 1 are
fitted allowing for a linear combination of all our five templates, also
for the cases of removing the outliers ({\it thick line}) or not
removing them ({\it thin line}).
Note that the thin dashed line is not exactly the same one as the thick
solid line in Figure \ref{fig:f10}: the reason is that here, the 68
percentile error is found over each redshift interval of width
$\Delta z = 0.1$, in the same way as for the SDSS galaxies, whereas in
previous plots the 68 percentile error is found first in bins of width
$\Delta z = 0.002$ and then averaged in quadrature over the larger width
$\Delta z = 0.1$. The errors are slightly reduced for this reason in
Figure \ref{fig:f16}, and in reality one would have to deal in some way
with the large oscillations with redshift shown in Figure \ref{fig:f7}, which 
are the main reason for the change in the error with the redshift
bin width used for extracting the 68 percentile value.
Even after the outliers are removed, the redshift errors obtained when
fitting the realistic LRG population are substantially increased
relative to fitting an ideal population matching template 1 exactly.
An error below $0.003 (1+z)$ is obtained only at $z < 0.55$
for $L=L_*$. The rms error is still larger, because even after removing
the outliers outside the dotted lines of Figure \ref{fig:f15}, the
error distribution is still substantially non-Gaussian.
Figure \ref{fig:f16} also shows that the number of outliers is much
smaller for galaxies generated with template 1. Real galaxy variability
introduces complications in fitting photometric redshifts that increase
the fraction of outliers.

  Overall, the 68 percentile redshift error for $L_*$ galaxies increases
by about 20\% for SDSS galaxies compared to template 1 galaxies after the
outliers are removed, and by $\sim 40$\% if the outliers are not removed.
Comparing also to the results when template 1 galaxies are fitted with
the same template 1 in Figure \ref{fig:f10}, we find that the redshift errors for the SDSS 
galaxies are 50\% to 60\% larger than the errors found when assuming that there is no 
galaxy variability at all, after removing the outliers, and a factor of 2 larger when outliers 
are not removed.

\begin{figure}
\includegraphics[width=\textwidth, angle=0]{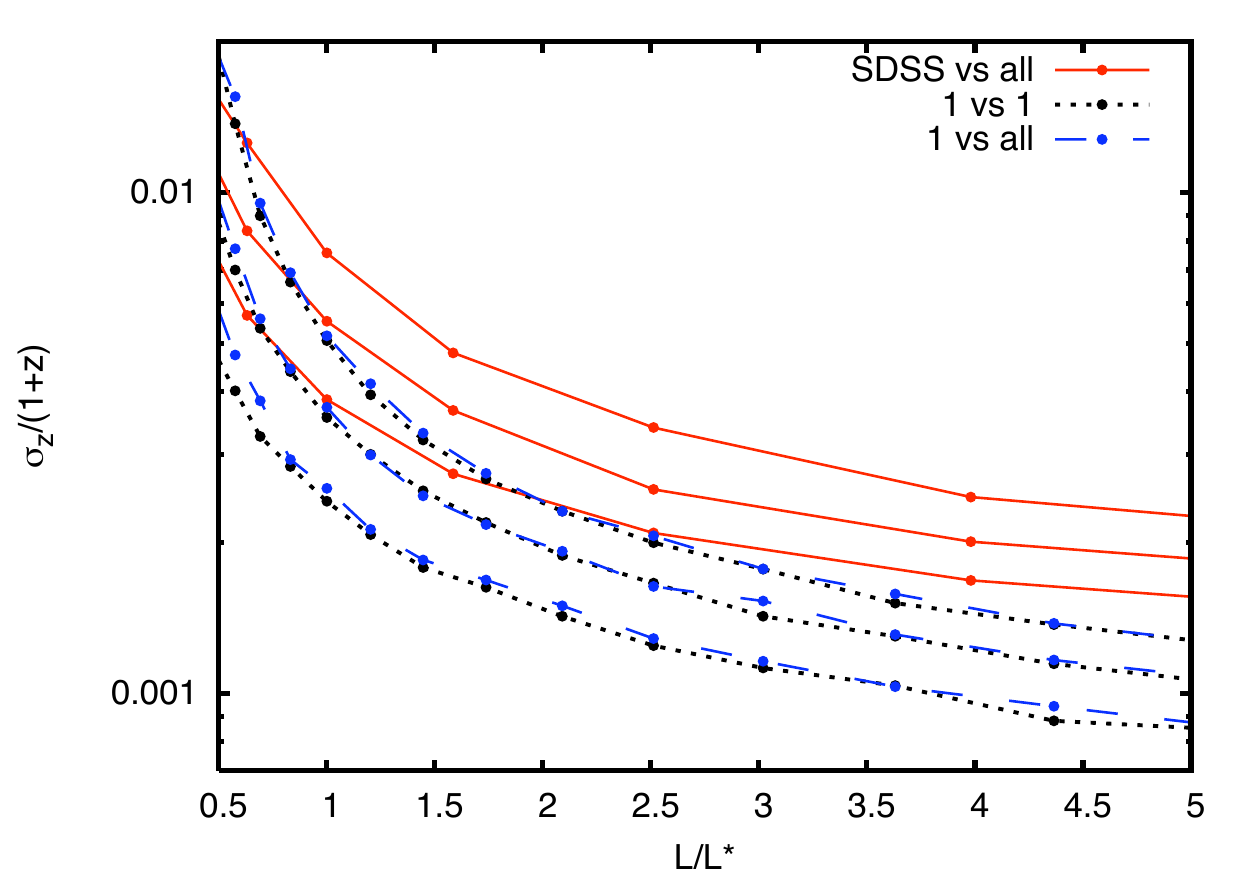}
\caption{Photo-z error as a function of luminosity.
{\it Solid lines:} Redshifted SDSS DR6 galaxies are fitted 
with our five templates. {\it Dashed lines:} Same results when galaxies
are generated with template 1, and fitted with five templates.
{\it Dotted lines:} Same results for galaxies generated with template 1
and fitted with the same template.
The redshift error is the 68 percentile of the averaged distribution
over each redshift interval of width $\Delta z = 0.1$, centered at
$z=0.65$, 0.75 and 0.85 (errors increase with redshift).}
\label{fig:flum}
\end{figure}

  The redshift errors after removing outliers are also shown in Figure
\ref{fig:flum}, as a function of galaxy luminosity, for several
redshifts. Results are presented for the SDSS galaxies and for the
cases of fitting template 1 galaxies with either template 1 alone or
all our five templates. The error in the SDSS galaxies decreases much
more slowly with luminosity than for mock galaxies generated with the
templates. This is simply because the variability of SDSS LRG spectra
is not well represented by our templates. For $L \gtrsim 2L_*$, 
the errors for the SDSS galaxies are nearly twice as large as the errors 
obtained from galaxies that match template 1 exactly, while the errors
when fitting with one or all five templates become nearly equal. The
fraction of outliers is also found to be much larger for the SDSS
galaxies.
The galaxy variability in the SDSS LRG is highly complex and cannot be
easily accounted for with a small number of templates, even when these
templates have been calibrated with observed galaxies.

We have checked the dependence of the redshift errors on the signal-to-noise 
reached in the photometry. For $N_f=30$, we find
that the photo-z error versus the signal-to-noise in the reddest filter
($S/N_{last}$) is quite robust to the various assumptions made about the
survey design. The target accuracy of $\sigma_{lz} = 0.003$ is reached
for SDSS galaxies at $S/N_{last} \simeq 20$, and at $S/N_{last} \simeq
12$ in the one template anlaysis. The photo-z error and the fraction of outliers
increase rapidly for lower signal-to-noise. At higher signal-to-noise,
the errors continue to decrease rapidly for galaxies generated with the
templates, but they flatten out to an "error-floor" of about
$\sigma_{lz}=0.002$ for the SDSS galaxies, the ``knee" being at around
$S/N_{last}\sim 30$, indicating an intrinsic galaxy variability that is
not adequately modelled by a small number of templates. 

  A possible problem when using the SDSS spectra of LRG to model
photometry might arise from spectroscopic calibration errors, which
could result in artificial variations of the simulated fluxes that are
not due to real galaxy variability but to calibration errors. In fact,
\cite{Benitez} argued that SDSS spectra cannot be used to
model variability for this reason; we reach a different conclusion. 
Even though the spectroscopic calibration errors are found to be
typically at the level of $\sim$ 4\% on wavelength scales of 100 nm when
comparing photometry and spectroscopy of stars (see
http://www.sdss.org/dr6/products/spectra/spectrophotometry.html ),
the precision of the photometric redshifts depends on measuring spectral
features at the smallest wavelength
scale allowed by the narrow-band filters (for LRG, this is mostly the
H$\alpha$ break). For a filter width of 10nm, the derived photometric
redshifts might be affected by spectroscopic calibration errors on a 
similar wavelength scale. At this scale, the
calibration errors are much smaller: they are certainly less than 2\%
from the analysis of quasar spectra in regions with no Ly$\alpha$
forest absorption (P. McDonald, private communication; see also Figure 
19 in \cite{McDonaldlymanalpha}, where the contribution to the flux variance
due to calibration errors from different scales in quasar spectra is
shown as a ratio to
the Ly${\alpha}$ forest power), and they should be less than 1\% from the most
recent calibration made with low-metallicity F-subdwarfs (D. Schlegel,
private communication).
We therefore conclude that the variations in the SDSS
spectra of LRG galaxies which result in increased photometric redshift
errors are mostly real and not due to spectroscopic calibration errors.

We emphasize that, owing to galaxy evolution, the variability present
in the SDSS spectra is a lower limit for the variability that should be
expected for LRG at higher redshift, as discussed in \S 5.4. 
The errors might possibly
be reduced by further optimization of the templates as a function of
galaxy luminosity and environment if galaxy variability at high redshift
were better understood, but how substantial an improvement may be
achievable in this way remains to be proven. 

\begin{figure}
\includegraphics[width=\columnwidth, angle=0]{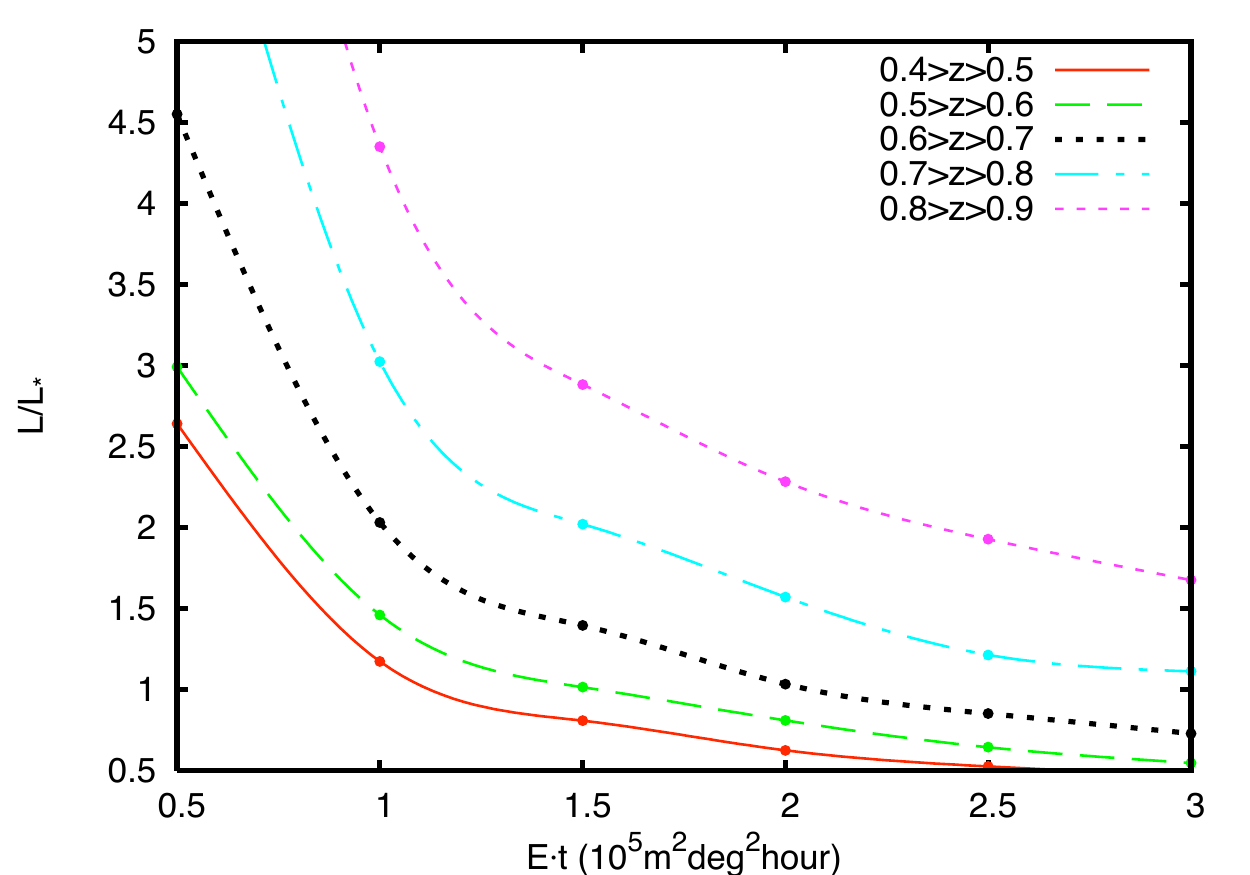}
\includegraphics[width=\columnwidth, angle=0]{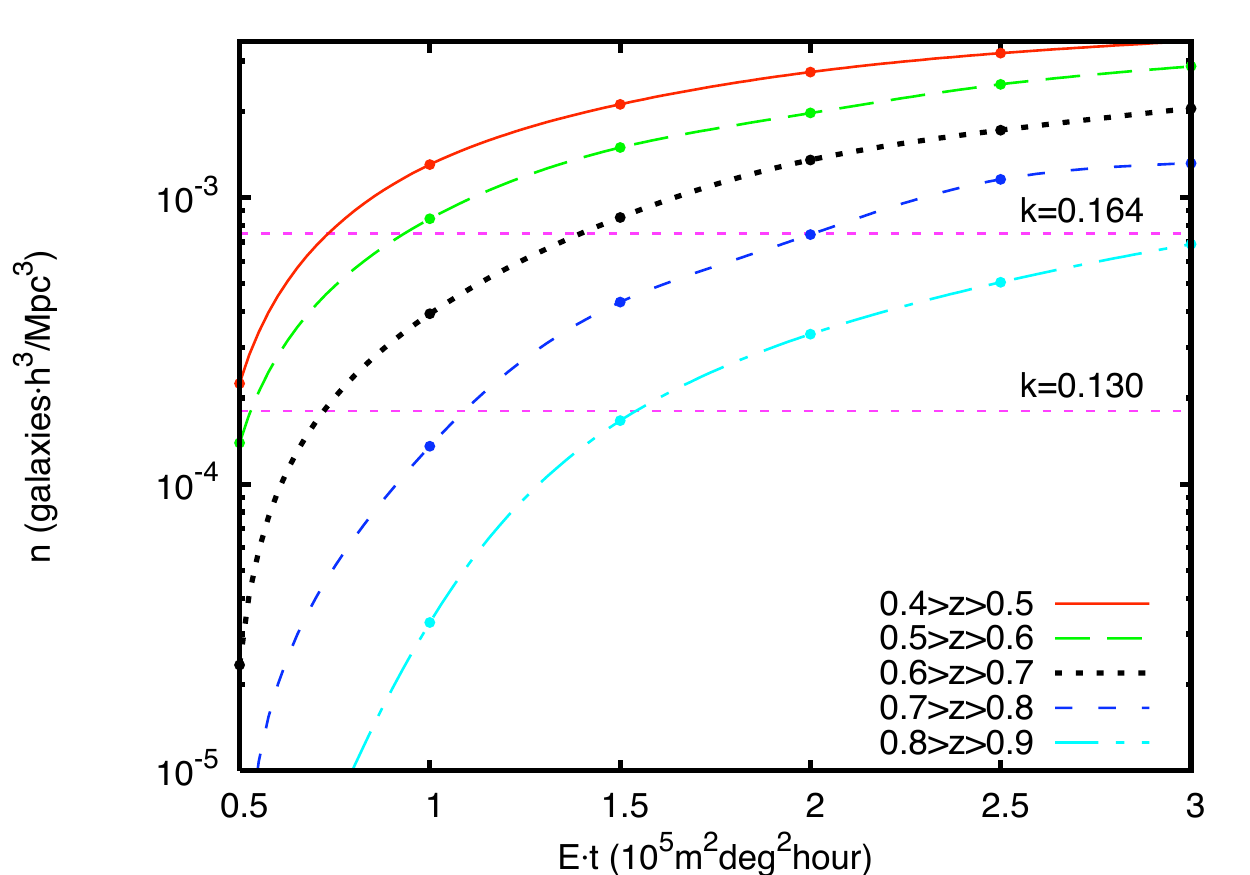}
\caption{{\it Top panel:} Luminosity of an LRG with photometric redshift
error $\sigma_{lz}=0.003$ after removing outliers, as a function of
the etendue $\times$ observing time of the survey, for the five redshift
intervals indicated. Errors are computed using the spectra of SDSS
LRG. {\it Bottom panel:} Corresponding number density of galaxies
with a photometric redshift error better than $0.003 (1+z)$, as a
function of $Et$, from the luminosity function \cite{Brown} after
taking out the fraction of outliers. Horizontal dashed lines indicate
the required galaxy densities for reaching a shot noise with $nP=1$ at
two wavenumbers (differing by a factor of two in the total number of
modes they comprise), where $P$ is the line-of-sight power spectrum
assuming $\sigma_{lz}=0.003 (1+z)$ and shown in Figure \ref{fig:pk}. 
Our  fiducial survey has $Et=1.5\times 10^{5}$ m$^2$ deg$^2$ hr.}
\label{fig:etendue}
\end{figure}

\section{Discussion}

 This paper has presented an analysis of the expected photometric
redshift errors in LRG that can be achieved in an imaging survey with
a system of a large number of narrow-band filters. We have found that an
optimal choice for the number of filters is $N_f \simeq 30$ over the
wavelength range $530$ to $830$ nm, allowing redshift measurements over
the range $0.5 < z < 0.9$. A central focus of our
investigation is that the intrinsic variability of LRG implies a
substantial increase of the redshift errors compared to a calculation
where all the LRG are assumed to match exactly the same template used
to fit their photometry. We find that it is crucial to properly include
the effects of spectral variability when evaluating the prospects for
any survey to measure BAO (particularly in the radial direction) in
the power spectrum of LRG.

 The implications of our results can be summarized by considering, for
any particular survey, the lowest luminosity that a galaxy needs to have
to yield a redshift accuracy better than the target level for measuring
BAO in the radial direction, $\sigma_{lz} < 0.003$. The power of a
survey to reach the required photometric accuracy depends mainly on the
product $Et$ of the etendue times the total observing time. This lowest
luminosity as a function of $Et$ is shown in the upper panel of Figure \ref{fig:etendue}, at 
five different redshift
intervals, when the total survey area is fixed to $\Omega = 8000$ square
degrees and $N_f=30$, for the SDSS galaxies after removing outliers with
redshift errors
$\Delta z > 0.05 (1+z)$. Note that the minimum luminosity shown at each
value of $Et$ is the one that would yield a 68 percentile error equal to
$\sigma_{lz}=0.003$; the rms error would be larger because even after
removing the most extreme outliers, the error distribution is still
substantially non-Gaussian. The luminosity function \cite{Brown} 
implies the number density of galaxies above this minimum luminosity
that is shown in the bottom panel, after subtracting outliers.
As shown in Figure \ref{fig:pk}, the photometric redshift errors suppress the power spectrum 
near the line-of-sight, which must be compensated by an increased number density of galaxies 
to keep shot noise sub-dominant on a given scale. The bottom panel of Figure \ref{fig:etendue} 
addresses this issue. For redshift errors $\sigma_{lz} = 0.003$, the two horizontal dashed lines  
show the galaxy number density required to reduce the  shot noise to the level $nP = 1$ at wavenumbers  
$k = 0.13$ $h$ Mpc$^{-1}$ and $k = 0.164$ $h$ Mpc$^{-1}$ near the line-of-sight. 
Modes up to $k = 0.164$ $h$ Mpc$^{-1}$  must be  sampled  to fully measure the second BAO.
Figure \ref{fig:etendue} shows that $Et > 3 \times  10^5$ m$^2$ deg$^{2}$ hr is required to
reach $nP = 1$  up to $z=0.85$ for $k = 0.164$ $h$ Mpc$^{-1}$ near the line-of-sight.
For the characteristic total observing time of 5000 hours
assumed in our fiducial survey, this corresponds to  $E > 60$ m$^{2}$ deg$^{2}$.

This minimum value for $Et$ depends on several technical details of
the telescope-camera system and the observing conditions for the survey.
The assumptions made here for our fiducial survey were specified in
\S 4: a seeing of $0.8$'', read-out noise of  $7 \times 3$ electrons
per pixel (with a pixel size of 0.4''), the efficiency for the latest
available CCD's, optical losses of two mirror reflections only, and the
sky brightness of the Paranal observatory at an airmass of $1.2$. These
assumptions can be considered as highly optimistic, except perhaps the
assumption on the read-out noise which might be improved with the best
available technology. Changes in these assumptions imply corresponding
changes in the required value of $Et$ necessary to reach a fixed
signal-to-noise in the galaxy photometry. We can quantify this in the
following way:
\begin{itemize}
\item A seeing degradation from $0.8$'' to $1.2$'' implies a 20\%
decrease in the signal-to-noise for a fixed aperture, and therefore
requires a $40$\% increase in $Et$ ($d\ln Et/d \ln seeing'' \simeq 1$)
\item An increase by half a magnitude of the sky brightness (comparable
for example to the sky variation over a sun cycle or between ecliptic
plane and ecliptic pole) can be compensated by a increase in $Et$ of a
factor $1.6$ ($d\ln Et/ dB_{sky}\simeq 0.9$)
\item A reduction in readout noise of a factor of three can be
compensated by a $\sim 20$\% reduction in $Et$
( $d \ln Et/d \ln N_{readout}= 0.3$; we have verified this by repeating
all the above calculations for a lower read-out noise).
An even lower read-out noise, however, would not yield the same
fractional improvement because at that point the read-out noise becomes
subdominant.
\item Any factors that decrease the overall throughput of the system
(i.e., lower efficiency of the CCD's, additional optical losses in the
telescope-camera system, reduced filter transmission, etc.) will of
course increase the required $Et$ in proportion to the inverse of the
throughput. An important factor that may cause such a reduction in the
throughput would likely be present in a system of narrow filters based
on interferometry, where the filter window has a variation across the
field-of-view caused by the varying incidence angle of the light. The
averaging of the filter window shape as a source moves across the field
in drift-scanning mode would widen the filter window without increasing
the number of detected photons.
\end{itemize}
These considerations mean that, in practice, the required etendue
would likely be substantially larger.

\section{Conclusions}

The principal objective of this paper is to examine the potential for narrow-band
photometric surveys to measure the radial  BAO signal  in the LRG  power spectrum. 
We have revisited this issue after \cite{SeoEisen03,SeoEisenstein07,Bridle}, addressing the
optimization of the number of filters and filter width, discussing the
dependence on the spectra of the target population of LRG, and
considering the signal-to-noise that could be realistically achieved in
a survey. We have found it to be particularly important to take into
account the spectral variability of a realistic galaxy population for
properly evaluating the photometric redshift accuracy that can be
achieved. Our conclusions can be summarized as follows:

{\it a)} In agreement with \cite{SeoEisen03, SeoEisenstein07, Bridle} the photometric approach, even if it can achieve the 
target photo-z error of 0.3\%, will be advantegeous only if it can cover a much larger fraction of the sky, with a higher galaxy 
density and reach higher redshifts than  the spectroscopic surveys currently under way. Not only one needs higher source density 
in a photometric survey to measure the oscillations up to $k \sim 0.2$, but also the distribution of redshift errors needs to be known 
accurately to correct for their effect on the power spectrum shape and be able to measure the BAO shape. This is equivalent to  the 
requirement of obtaining photo-z  with better than 0.3\% accuracy to fainter magnitudes than spectroscopy.
We have  concentrated on the LRG galaxies  because they are a fairly homogeneous population,  the form of their spectra makes 
them particularly suitable for  good photo-z determinations and because they represent the target of choice for $z<2$  BAO surveys.

{\it b)} If LRG were a perfectly homogeneous population (with spectra
closely matched by a single spectral template), the photometric redshift
accuracy would not improve very much beyond a total number of filters
$N_f \simeq 10$, except for galaxies at high signal-to-noise which would
have a low number density in a survey with $Et\simeq 1.5 \times 10^5\,
{\rm m}^2\, {\rm deg}^2$ hr. In reality, galaxy spectra are variable and
the ideal number of filters is $N_f\simeq 30$ over a wavelength range
$530\, {\rm nm} < \lambda < 830\, {\rm nm}$, focusing on the redshift
range $0.5 < z < 0.9$. Medium-to-narrow band filters photometric  redshift errors  
show periodic oscillations with a spacing $\delta z$ that corresponds to the filter 
widths. For a  filter width of $\sim 100 \AA$,  the scale of the oscillation corresponds 
to the BAO scale. The presence of these oscillations may introduce significant biases 
in the measurement of the BAO scale (see Section 5.2).

{\it c)} The variability of realistic galaxy spectra demand high resolution
(or more filters in a photometric survey) and high signal-to-noise.  We have 
quantified, with realistic simulations of a photometric survey, the  minimum 
galaxy luminosity and the corresponding number density for which 0.3\% photo-z error  
can be obtained as a function of the survey Etendue$\times$  exposure time $Et$ (see Figure \ref{fig:etendue}). 
We estimate the minimum value of $Et$ necessary to
measure the full second baryonic acoustic oscillation in the power
spectrum, at $k<0.164$ $h$/Mpc  near the line-of-sight, with shot
noise $nP > 1$ and up to $z=0.85$, at
$Et\simeq 3\times 10^5\, {\rm m}^2\, {\rm deg}^2$ hr. The actual required
value of $Et$ would likely be larger when taking into account that our 
assumptions for the seeing and the overall throughput of the
telescope-camera system are optimistic, and that the presence of outliers, 
optimistically clipped and ignored, will degrade the estimated performance. Moreover, the true level
of galaxy variability should likely be higher than in the SDSS LRG
sample we have used (which is for $z \lesssim 0.4$ and $L \gtrsim 3
L_*$), because the variability of LRG spectra is expected to increase
with redshift and decrease with luminosity.

Current spectroscopic surveys (e.g.,  SDSSIII--BOSS) are obtaining spectra of the massive ($ > 2 L^{*}$) 
population of LRG at $z<0.75$ to measure radial and tangential BAO scale. Here we have investigated the 
conditions necessary to obtain the required photo-z accuracy of $0.3$\% from a similar or  fainter population 
using photometry, which in principle is less expensive in terms of observing. We have concluded that  an 
etendue of $30$ ${\rm m}^2\, {\rm deg}^2$ is required for tracing the same population as BOSS ($z<0.75$, $L > 2 L^{*}$)  with radial 
BAO signal degraded  only by $\sim 60$\% due to photo-z errors. But to improve on this i.e.,  to reach  
$L_*$ galaxies   at slightly higher redshifts,  the minimum etendue required is $60$ ${\rm m}^2\, {\rm deg}^2$ for a 
5-year survey (corresponding to a 4m telescope with the largest fields of view that have been made), and probably 
needs to be larger given our optimistic assumptions we have made for the seeing, throughput,
observing time, removal of outliers and systematic errors.

In this paper we have restricted our attention to the objective of measuring radial BAO
with LRG. However, a large-area imaging optical survey with a large number of
narrower bands than existing surveys would have many other applications, and might lead to a large
number of interesting astronomical discoveries. 
We have shown in this  paper that, for realistically achievable values of $Et$ at the present time, the 
density  of LRG sources that could be measured with a narrow-band photometric survey with 
the target redshift accuracy for radial BAO will not be large enough
to make it competitive with a spectroscopic survey. Therefore, a photometric survey 
with a large number of optical narrow-bands needs to find its scientific justification in 
other astronomical applications.

\section*{Acknowledgements}
We warmly thank D. Eisenstein, J. Gunn, P. McDonald and D. Schlegel for
discussions. 
This work was carried out in the framework of the PAU Consolider
Collaboration: the authors are part of the Physics of the Accelerating
Universe (PAU) proposal, currently supported by the Spanish Ministry
for Science and Innovation (MICINN) through the Consolider Ingenio-2010
program project CSD2007-00060.
DR acknowledges support from the Spanish MICINN through a FPU grant.
LV acknowledges the support of FP7-PEOPLE-2002IRG4-4-IRG\#202182 and CSIC I3 \#200750I034. The work of RJ is supported by grants from the Spanish MICINN and the European Union (FP7). JM is supported by the Spanish MICINN grants AYA2006-06341 and AYA-15623-C02-01 and the European Union FP6 grant IRG-046435. CPG acknowledges MICINN grant No. FPA2007-60323.

\section*{References}
\bibliographystyle{JHEP}
\providecommand{\href}[2]{#2}\begingroup\raggedright

\end{document}